\documentclass[prd, onecolumn, superscriptaddress, nofootinbib, notitlepage, floatfix]{revtex4-1}
\usepackage{latexsym,amsmath,amssymb,amstext}
\usepackage{slashed}
\usepackage{epsfig}
\usepackage{float,xcolor}
\usepackage{graphicx}
\usepackage{dcolumn}
\usepackage{comment}

\newcommand{\be}{\begin{equation}}
\newcommand{\ee}{\end{equation}}
\newcommand{\bea}{\begin{eqnarray}}
\newcommand{\eea}{\end{eqnarray}}
\newcommand\bef{\begin{figure*}}
\newcommand\eef[1]{\label{fg:#1}\end{figure*}}
\newcommand\beq{\begin{equation}}
\newcommand\eeq[1]{\label{#1}\end{equation}}
\newcommand\beqa{\begin{eqnarray}}
\newcommand\eeqa[1]{\label{#1}\end{eqnarray}}
\newcommand\bet{\begin{table}}
\newcommand\eet[1]{\label{tb:#1}\end{table}}

\newcommand\fgn[1]{Figure \ref{fg:#1}}
\newcommand\eqn[1]{Eq.\ (\ref{#1})}

\newcommand{\lat}{\mathrm{lat}}
\newcommand{\phys}{\mathrm{phys}}

\begin{document}

\widetext

\title{Scaling dimension of $4\pi$-flux monopole operator  in
four-flavor three-dimensional QED using lattice simulation}

\author{Nikhil\ \surname{Karthik}}
\email{nkarthik.work@gmail.com}
\affiliation{American Physical Society, Hauppauge, New York 11788}
\affiliation{Department of Physics, Florida International University, Miami, FL 33199}
\author{Rajamani\ \surname{Narayanan}}
\email{rajamani.narayanan@fiu.edu}
\affiliation{Department of Physics, Florida International University, Miami, FL 33199}

\begin{abstract}

We numerically address the issue of which monopole operators are
relevant under renormalization group flow in three-dimensional
parity-invariant noncompact QED with $4$ flavors of massless
two-component Dirac fermion.  Using lattice simulation and finite-size
scaling analysis of the free energy to introduce monopole-antimonopole
pairs in  $N=4$ and $N=12$ flavor noncompact QED$_3$, we estimate
the infrared scaling dimensions of monopole operators that introduce
$2\pi$ and $4\pi$ fluxes around them.  We first show that the
estimates for the monopole scaling dimensions are consistent with
the large-$N$ expectations for $N=12$ QED$_3$. Applying the same
procedure in $N=4$ QED$_3$, we estimate the scaling dimension of
$4\pi$ flux monopole operator to be $3.7(3)$, which allows the
possibility of the operator being irrelevant.  This finding offers
support to the scenario in which higher-flux monopoles are irrelevant
deformations to the Dirac spin liquid phase that could be realized
on certain non-bipartite lattices by forbidding $2\pi$-flux monopoles.

\end{abstract}

\maketitle

\section{Introduction}\label{sec:intro}

The characterization of the quantum numbers of monopoles in the
continuum and on various lattices, and their impact on the infrared
behavior of three-dimensional quantum electrodynamics coupled to
even number of flavors ($N$) of massless two-component Dirac fermions
have been topics of interest in recent times. The presence and
absence of monopoles are expected to change the long-distance
behavior of QED$_3$ radically. QED$_3$ without monopole excitation
-- noncompact QED$_3$ -- has been intensely studied using various
methods, for example, using lattice
regularization~\cite{Hands:1989mv,Hands:2002dv,Hands:2004bh,Raviv:2014xna},
conformal
bootstrap~\cite{Chester:2016wrc,Chester:2017vdh,He:2021sto,Li:2021emd,Albayrak:2021xtd,Rychkov:2023wsd},
Dyson-Schwinger
approaches~\cite{Pisarski:1984dj,Appelquist:1985vf,Appelquist:1986qw,Appelquist:1986fd,Appelquist:1988sr,Gusynin:2003ww,Gusynin:2016som,Kotikov:2016wrb}.
Ab initio numerical studies using the lattice regularization with
exactly massless fermions and finite-size scaling have shown that
parity-invariant non-compact QED$_3$ exhibits scale-invariant
behavior independent of the number of flavors; some salient
observations toward this conclusion stems from finite-size scaling
of low-lying Dirac eigenvalues~\cite{Karthik:2015sgq,Karthik:2016ppr}
and closer resemblance of their eigenvalue distributions to those
from a simple conformal model~\cite{Karthik:2020shl}, and presence
of power-law correlators~\cite{Karthik:2016ppr,Karthik:2017hol}.
Despite the term {\sl noncompact}, it should be emphasized that the
fermions see a compact version of the gauge field in the lattice
regularization.  On the other hand, QED$_3$ with any number of
monopoles -- compact QED$_3$ -- without massless fermion content
is well known to be confined~\cite{Polyakov:1975rs,Polyakov:1976fu}.

Even though monopoles do not arise dynamically in the noncompact
theory after the ultraviolet regulator is removed, one could subject
the noncompact theory to monopole-like singular boundary conditions
at various space-time points; for a flux $Q$ monopole, the total
flux~\footnote{An alternate convention is to count flux as $4\pi
q$ with $q=1/2,1,\ldots$ corresponding to our convention $Q=1,2,\ldots$}
on surfaces enclosing the point is $2\pi Q$ for integers $Q$.  For
fermions coupled to the U(1) gauge fields, the extended Dirac string
singularity is invisible, and the insertion of the monopole behaves
like the insertion of a composite operator at the point.  Hence,
one defines the monopole operator ${\cal M}_Q(x)$ through its action
of introducing  $2\pi Q$ flux around the point $x$~\cite{Borokhov:2002ib}.
At critical points of a U(1) lattice theory, one can find the scaling
dimension of such monopole operators via the two-point functions,
\beq
\left\langle {\cal M}^\dagger_Q(x) {\cal M}_Q(0)\right\rangle \propto |x|^{-2\Delta_Q}.
\eeq{mono2pt}
The exponent $\Delta_Q$ is the scaling dimension of ${\cal M}_Q$.
Since criticality is approached only in the long-distance or the infrared limit of
QED$_3$, the above power-law scaling will be seen when the monopole
and antimonopole are separated by large distances.  If the infrared
dimension $\Delta_Q >3$, then flux-$Q$ monopoles are irrelevant
to the infrared end of the renormalization group flow.

The nature of compact QED$_3$ coupled to exactly massless fermions
is yet unresolved.  A lattice study of compact QED$_3$ is complicated
by the inherent singular nature of monopoles~\cite{Karthik:2019jds}
and the presence of near-zero Dirac operator eigenmodes away from
the massless limit even as one decreases the lattice spacing. It
is conceivable that a naive continuum limit of compact QED$_3$ with
massless fermions in a traditional sense (that is, an approach to
continuum limit at the Gaussian fixed point as the lattice coupling
is taken to zero keeping physical scales fixed) is not well-defined.
Refs~\cite{Hands:2006dh,Armour:2011zx} studied this theory in the
presence of a four-fermi interaction. Another option might be to
UV complete compact QED$_3$ using a SU(2) theory in the presence
of a Higgs field~\cite{Polyakov:1976fu}. A quantum Monte-Carlo
study~\cite{Xu:2018wyg} instead focused on a U(1) lattice gauge
theory coupled to many flavors of staggered fermions and its
strong-to-weak coupling phase diagram. This fundamental difficulty
associated with a direct numerical study of the UV-IR renormalization
group flows in compact QED$_3$ in the continuum limit provides a
strong motivation to study the scaling dimension of monopole operators
in the noncompact version of the theory. In compact QED$_3$, monopoles
of all $2\pi Q$ fluxes dynamically appear.  As $\Delta_Q$ is typically
a monotonically increasing function of $Q$, if $\Delta_1$ of the
$Q=1$ monopole is greater than 3 in an $N$ flavor noncompact QED$_3$,
then one expects a similar conformal behavior in the $N$-flavor
compact QED$_3$ and noncompact QED$_3$. In this way, the $Q=1$
monopole is expected to be relevant only for $N<12$ based on large-$N$
and $4-\epsilon$ approximations~\cite{Pufu:2013vpa,Chester:2015wao},
and further confirmed by lattice simulations in previous as well
as in the present work. By the above argument, the compact QED$_3$
is expected to have a critical number of flavor $N\approx 12$.

Given the dominant role of the monopole creating the smallest $2\pi$
flux, a study of higher flux-creating monopoles might not seem
significant.  However, the specific motivation for studying monopoles
that create $4\pi$ flux in this work is the following.  Recently,
there has been interest in a realization of compact QED$_3$ where
the dominant $Q=1$ monopole is disallowed due to ultraviolet
symmetries specific to certain lattices~\cite{Song:2018ial}.  Such
a version of $N=4$ compact QED$_3$ that is devoid of $Q=1$ monopole
is expected to be an effective field theory description of the
antiferromagnetic Heisenberg spin model on Kagom\'e and triangular
lattices that could host a Dirac spin liquid (DSL)
phase~\cite{Song:2018ial,Song:2018ccm,Zhu:2018thc}.  Whereas the
stability of DSL on a triangular lattice is also decided by the
requirement of irrelevance of an allowed four-fermi term that is
close to being marginal, the $Q=2$ monopole was argued~\cite{Song:2018ccm}
to be the critical object determining the stability of DSL on
Kagom\'e lattices. At the present accuracy of the large-$N$ expansion
of QED$_3$, the scaling dimension of the $Q=2$ monopole operator
is approximately 2.5. Since this value is quite close to the marginal
value of 3, it is possible that the higher-order perturbative
corrections and genuine nonperturbative corrections to the large-$N$
value at the relatively small $N=4$ could shift the actual value
of the monopole scaling dimension to be greater than 3.  Therefore,
the possibility of the long-distance correlation in the DSL phase
on the two non-bipartite lattices, especially on the Kagom\'e
lattice, is then tied to the infrared conformality of the $N=4$
compact QED$_3$ in the absence of $Q=1$ monopoles; in other terms,
to the infrared irrelevance of the next-allowed $Q=2$ monopole
operators in $N=4$ noncompact QED$_3$.  The nonperturbative
determination of the scaling dimension of the $Q=2$ monopoles in
$N=4$ noncompact QED$_3$ using direct lattice simulation is therefore
the main aim of this paper.

\section{Method}\label{sec:method}

\subsection{Lattice regulated noncompact QED$_3$ and monopole insertions therein}

The parity-invariant noncompact theory consists of Abelian gauge
fields $A_\mu(x)$ coupled to an even number of flavors, $N$, of
massless two-component Dirac fermions.  The gauge coupling $g^2$
in the theory has a mass dimension of 1, which makes the theory
super-renormalizable and we can use appropriate factors of $g^2$
to make all masses and lengths dimensionless. We study the lattice
regulated version of the theory on a periodic box of dimensionless
physical volume $\ell^3$ that is discretized using a lattice of
size $L^3$.  The continuum limit of the finite volume theory can
be obtained by extrapolating to $L\to\infty$ limit in different
fixed physical extents $\ell$.

We take a brief detour to formally define noncompact gauge theory
that is explicitly a U(1) gauge theory, and define monopole insertions
within this U(1) lattice gauge theory.  The Villain
formulation~\cite{Villain:1974ir} of the noncompact QED$_3$ can be
defined via the path integral,
\beq
Z=  \left(\prod_{x,\mu} \int_{-\infty}^\infty d\theta_\mu(x)\right) \det{}^{N/2}\left[\slashed{C}\slashed{C}^\dagger\right] {\cal W}_g(\theta),
\eeq{latticepath}
where $\slashed{C}$ is a two-component lattice Dirac operator that
is coupled to the lattice gauge fields $\theta_\mu(x) = A_\mu(x)
\ell/L$ via the compact variable $U_\mu(x) = e^{i\theta_\mu(x)}$.
The theory is regulated in a parity-invariant manner by coupling
$N/2$ flavors  to $\slashed{C}$, and the other $N/2$ to
$\slashed{C}^\dagger$. The contribution from the gauge sector is
${\cal W}_g(\theta)$ given by,
\beq
{\cal W}_g(\theta) = \sum_{\{N_{\mu\nu}\}}\exp\left[-\frac{L}{\ell}\sum_{x}\sum_{\mu>\nu}\left(F_{\mu\nu}(x) - 2\pi N_{\mu\nu}(x)\right)^2\right]\quad\text{where}\quad
F_{\mu\nu}(x) = \Delta_\mu\theta_\nu(x)-\Delta_\nu\theta_\mu(x).
\eeq{wgdef}
with the sum over configurations of integer values $N_{\mu\nu}(x)$
associated with the $(\mu,\nu)$-plaquette at site $x$.  Since both
${\cal W}_g(\theta)$ and the Dirac operator $\slashed{C}$ are
invariant under shifts $\theta_\mu(x) \to \theta_\mu(x) + 2\pi
n_\mu(x)$ for integers $n_\mu(x)$, the path-integral \eqn{latticepath}
is that of U(1) gauge theory coupled to fermions.  The magnetic
charge $Q(x)$ of the monopole at a site $x$ is
defined~\cite{DeGrand:1980eq} via the divergence of an integer-valued
current dual to $N_{\mu\nu}(x)$; that is,
\beq
Q(x) \equiv \frac{1}{2}\sum_{\mu,\nu,\rho}\epsilon_{\mu\nu\rho}\Delta_{\mu}N_{\nu\rho}(x).
\eeq{monocharge}
For the sake of brevity, we simply define a monopole at a point $x$
with a value of net flux as $2\pi Q$ as a flux-$Q$ monopole.
Depending on the constraints on the allowed values of $Q(x)$ in the
path-integral, which thereby corresponds to constraints on the
allowed configurations $\{N_{\mu\nu}\}$ in \eqn{wgdef}, one can
define different versions of QED$_3$.  The noncompact QED$_3$ is
the U(1) gauge theory with the constraint, $Q(x)=0$, at all $x$ in
the continuum limit of the lattice-regulated theory. In this case,
by making use of the invariance of theory under $\theta_\mu \to
\theta_\mu + 2\pi n_\mu$ shifts, one can write down the path-integral
in the usual form without any sum over $N_{\mu\nu}$ as
\beq
Z_0 =  \left(\prod_{x,\mu} \int_{-\infty}^\infty d\theta_\mu(x)\right) \det{}^{N/2}\left[\slashed{C}\slashed{C}^\dagger\right] e^{-\frac{L}{\ell} \sum_{x}\sum_{\mu>\nu}F_{\mu\nu}(x)^2}.
\eeq{normalpath}
We can define the path-integral $Z_Q$ with a flux-$Q$ monopole inserted at a point $x'$ and an antimonopole at a point $x'+r$ by subjecting to the constraint $N_{\mu\nu}(x)=N^{Q\bar Q}_{\mu\nu}(x;r)$ with
\beq
\frac{1}{2}\sum_{\mu,\nu,\rho}\epsilon_{\mu\nu\rho}\Delta_{\mu} N^{Q\bar Q}_{\nu\rho}(x;r) = Q \delta_{x,x'} - Q \delta_{x,x'+r}.
\eeq{Nconstraint}
The two-point function of a monopole at $x'$ and $x'+r$ is simply
the ratio, $Z_Q/Z_0$.

The universal aspects like the anomalous dimensions at the infrared
fixed point should not be sensitive to the exact details of the
lattice operator so long as the operator quantum numbers are captured
correctly. Therefore, we can choose the type of background flux to
better capture the effect of monopole operators.  Instead of
introducing integer-valued flux $N^{Q\bar Q}_{\mu\nu}$ in the
path-integral, we follow the approach of
Refs~\cite{Murthy:1989ps,Pufu:2013eda} to introduce a classical
background gauge field ${\cal A}^{Q\bar Q}_{\mu}(x;r)$ that minimizes
the pure gauge action
\beq
S_g = \left[ B_{\mu\nu}^{Q\bar Q}(x;r) - 2\pi N^{Q\bar Q}_{\mu\nu}(x;r)\right]^2; \qquad B_{\mu\nu}^{Q\bar Q}(x;r) = \Delta_\mu {\cal A}^{Q\bar Q}_{\nu}(x;r) - \Delta_\nu {\cal A}^{Q\bar Q}_{\mu}(x;r),
\eeq{bmunu}
on a periodic box.
We then define the path-integral in the presence of monopole insertions via,
\beq
Z_Q =  \left(\prod_{x,\mu} \int_{-\infty}^\infty d\theta_\mu(x)\right) \det{}^{N/2}\left[\slashed{C}\slashed{C}^\dagger\right] e^{-\frac{L}{\ell} \sum_{x}\sum_{\mu>\nu}\left(F_{\mu\nu}(x)- Q B^{1\bar 1}_{\mu\nu}(x;r)\right)^2},
\eeq{normalpathQ}
using the fact that $B_{\mu\nu}^{Q\bar Q} = Q B_{\mu\nu}^{1\bar
1}$.  The above procedure has the advantage that the effect of
monopole can be completely removed from the path-integral in the
pure-gauge theory ($N=0$) by redefining the dynamical gauge field
$\theta_\mu(x) \to \theta_\mu(x) - {\cal A}^{Q\bar Q}_\mu(x;r)$.
Therefore any non-zero effect of the monopole at finite non-zero
$N$ can arise only due to the presence of massless fermions.  This
procedure has been put to test previously in the free fermion
theory~\cite{Karthik:2018rcg}, and at the critical point of the 3d
XY-model~\cite{Karthik:2018rcg}.

\subsection{Monopole correlator and finite-size scaling}

The lattice monopole two-point function is given by 
\beq
G_\lat(r,\ell,L) = \frac{Z_Q}{Z_0}.
\eeq{baretwopoint}
As with correlators of regular composite operators composed of local
fields, we assume that the lattice correlator $G_\lat$, which is
in units of lattice spacing $a=\ell/L$, can be converted into a
correlator in physical units $G_\phys$ by a multiplicative factor;
namely,
\beq
G_\phys(r,\ell,L) = a^{-2D_Q^{(N)}} G_\lat(r,\ell,L),
\eeq{renormalization}
where $D_Q^{(N)}$ is the ultraviolet exponent governing the monopole
correlator at short distances.  We will discuss more on the conversion
from the lattice to the physical correlator when presenting the
results from our numerical calculation.  The physical continuum
correlator, $G_\phys(r,\ell)$, after extrapolating to $L\to\infty$ will
show scale-invariant behavior at large separations $|r|$ and $\ell$
as
\beq
G_\phys(r,\ell) = \frac{1}{|r|^{2\Delta^{(N)}_Q}} {\cal G}\left(\frac{|r|}{\ell}\right)\quad\text{as}\quad |r|\to\infty,
\eeq{scaleinvcorr}
for some scaling function ${\cal G}$.  The exponent $\Delta^{(N)}_Q$
that governs the long-distance correlator is the infrared scaling
dimension of the monopole operator that we are seeking.  There could
be corrections to the above simple scaling from higher-order $1/\ell$
corrections and due to contamination from higher-dimensional flux-$Q$
monopole operators that the background field method could overlap
with; we assume such corrections are much smaller compared to the
numerical accuracy of our data.  By keeping the ratio $|r|/\ell =
\rho$ fixed as $\ell$ is increased,
\beq
G_\phys(|r|=\rho\ell,\ell) \propto \frac{1}{\ell^{2\Delta^{(N)}_Q}}\quad\text{as}\quad \ell\to\infty.
\eeq{scaleinvcorr2}
We will follow this procedure in this work, and keep the
monopole-antimonopole separation proportional to box size, thereby
reducing the determination of the infrared scaling dimension to a
finite-size scaling analysis.

\bef
\centering
\includegraphics[scale=0.6]{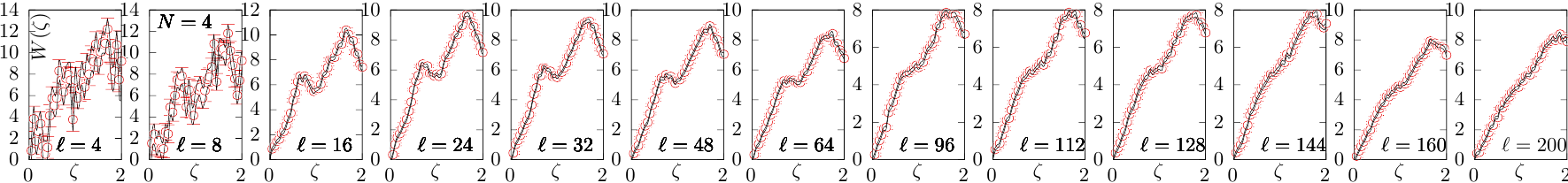}
\includegraphics[scale=0.6]{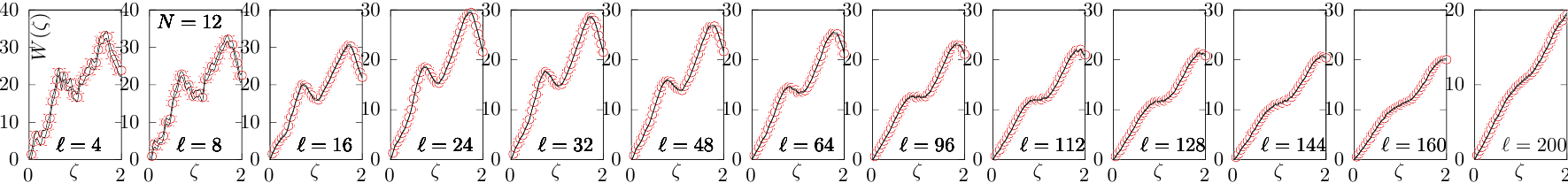}
\caption{Representative data points and their interpolations for the lattice
determined derivative of free energy, $W(\zeta)$, with respect to
the auxiliary parameter $\zeta$. The top and bottom panels show
results for  $W(\zeta)$ at different $\ell$ on $L=24$ lattice in
$N=4$ and 12 noncompact QED$_3$ respectively. The red data points
are the actual Monte Carlo determinations. The black bands are the
spline interpolations to the data. }
\eef{wversusz}

\subsection{Implementation of the numerical calculation}

We studied noncompact theory with $N=4$ and $N=12$ flavors on
periodic Euclidean boxes at multiple values of physical extents
$\ell=$ 4, 8, 16, 24, 32, 48, 64, 96, 128, 144, 160 and 200.  We
discretized them on lattices of volume $L^3$ with $L=12, 16,20,24$
and 28. We used Wilson-Dirac operator $\slashed{C}_w$ that is coupled
to 1-step HYP-smeared gauge field. We tuned to the massless point
by tuning the bare Wilson fermion mass $m_w$ so that the first
eigenvalue $\Lambda_1^2(m_w)$ of $\slashed{C}_w\slashed{C}_w^\dagger$
is minimized as a function of $m_w$. More details on the two-component
Wilson-Dirac operator and its mass tuning can be in our earlier
work in Ref~\cite{Karthik:2015sgq}.

We chose the displacement vector $r$ between the flux-$Q$ monopole
and antimonopole to be along one of the axis; namely, the three-vector
$r=(0,0,t )$ for $t=aT$ and integers $T$.  We kept $\rho = t/\ell=1/4$,
an arbitrary choice in the work to simplify the analysis to a
finite-size scaling one as explained before.  For this choice of
on-axis $r$, a natural choice for $N_{\mu\nu}^{1\bar 1}$ that
satisfies \eqn{Nconstraint} is $N_{12}^{1\bar 1}(0,0,x_3) = 1$ for
$x_3\in[1,T]$, and all other $N_{\mu\nu}^{1\bar 1}$ are set to 0.
Such a choice can be changed arbitrarily by shifts $N_{\mu\nu}^{1\bar
1} \to N_{\mu\nu}^{1\bar 1} + \Delta_\mu n_\nu - \Delta_\nu n_\mu$
for integers $n_\mu(x)$, that move and bend the Dirac string (the
column of plaquettes with $2\pi$ flux) keeping the location of
monopole and antimonopole fixed; however such variations are
unimportant in the U(1) theory, and therefore, the simplest choice
above for $N_{12}^{Q\bar Q}$ suffices. With this choice of
$N_{\mu\nu}^{1\bar 1}$, we determined the background field ${\cal
A}^{1\bar 1}_\mu(x)$, and the field tensor $B^{1\bar 1}_{\mu\nu}(x)$
in the periodic $L^3$ box by analytically minimizing~\cite{Karthik:2019jds}
the action \eqn{bmunu}. From this, the background field for any
value of $Q$ can be obtained as $Q B^{1\bar 1}_{\mu\nu}(x)$.

\bef
\centering
\includegraphics[scale=0.55]{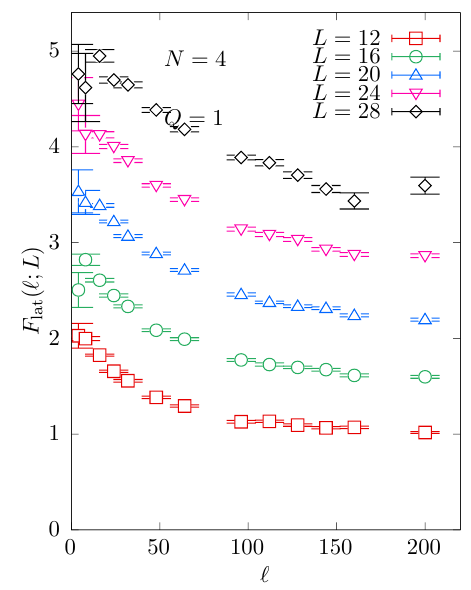}
\includegraphics[scale=0.55]{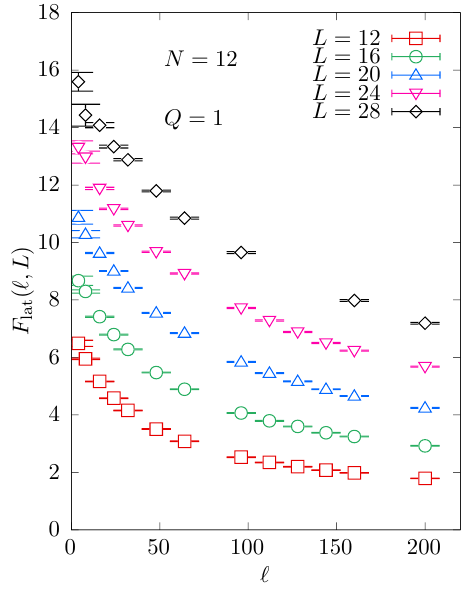}
\includegraphics[scale=0.55]{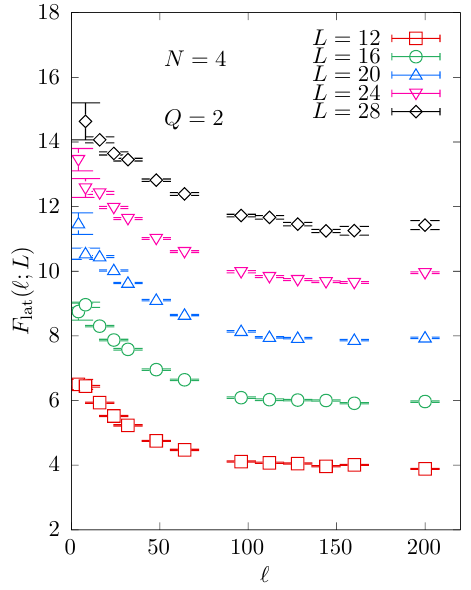}
\includegraphics[scale=0.55]{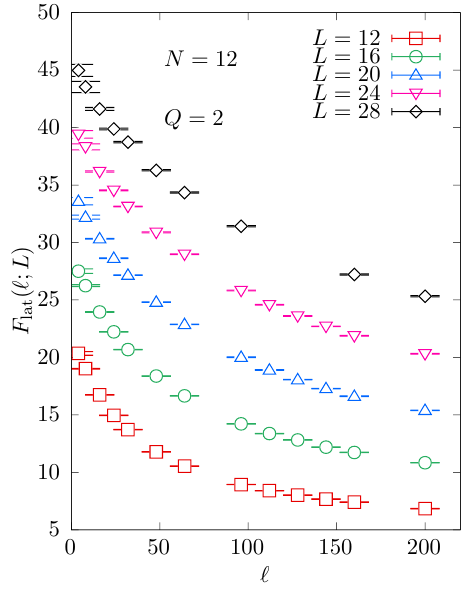}

\caption{The free energy, $F_\lat(\ell,L)$, in lattice units is
shown as a function of physical box size, $\ell$.  The values of
$(Q,N)$ in the panels from left to right correspond to $(1,4),
(1,12), (2,4), (2,12)$ respectively. In each panel, the data at
$L=12,16,20,24$ and 28 are shown.
}
\eef{fbare}

The effect of $B^{Q\bar Q}_{\mu\nu}$ is exponentially suppressed
in \eqn{normalpathQ}, and therefore, it is hard to compute $Z_Q$
as an expectation value in $Q=0$ theory.  Instead, we follow the
approach in Ref~\cite{Karthik:2018rcg}, and computed the logarithm
of the above correlator, which is nothing but the free energy in
lattice units to introduce a monopole-antimonopole pair, as
\beq
F_\lat(r,\ell,L)\equiv -\log(G_\lat(r,\ell,L)) = \int_0^Q W(\zeta) d\zeta,
\eeq{barefreeenergy}
where 
\beq
W(\zeta) = -\frac{1}{Z_\zeta}\frac{\partial}{\partial\zeta}Z_{\zeta},
\eeq{wdef}
and $Z_\zeta$ is the extension of the path-integral in \eqn{normalpathQ}
by the replacement $Q\to \zeta$ for real values of $\zeta$.  We
have simply differentiated $F_\lat$ with respect to an auxiliary
variable $\zeta$ and integrated it back again. The reason behind
doing so is that the quantity $W(\zeta)$ is computable as expectation
values, $\langle\cdots\rangle_\zeta$, in the Monte Carlo simulation
of the $Z_\zeta$ path-integral; namely,
\beq
W(\zeta) = \frac{2L}{\ell}\sum_{\mu>\nu}\sum_{x} B_{\mu\nu}^{1\bar 1}(x;r) \left\langle F_{\mu\nu}(x) - \zeta B_{\mu\nu}^{1\bar 1}(x;r)\right\rangle_\zeta.
\eeq{wzetameas}
We used 40 different equally spaced values of $\zeta\in[0,2]$.  At
each value of $\zeta$, we performed independent hybrid Monte Carlo
(HMC) simulation of $Z_\zeta$ to compute $W(\zeta)$ numerically.
From each thermalized HMC run, we generated between 15K to 30K
measurements of $W(\zeta)$. By using Jack-knife analysis, we took
care of autocorrelations in the collected measurements.

\section{Results}\label{sec:results}

\subsection{Determination of free energy}

From the Monte Carlo simulation, we collected the data for $W(\zeta)$
from the relation in \eqn{wzetameas}.  In \fgn{wversusz}, we show
the numerically determined $W(\zeta)$ (the red circles in the panels)
as a function of $\zeta$ at all $\ell$ on a fixed $L=24$ lattice.
We show the data from $N=4$ and 12 flavor theories in the set of
top and bottom panels respectively. The actual simulation points
span $\zeta\in[0,2]$.  In order to perform the needed integration
in \eqn{barefreeenergy}, we interpolated the data between 0 and 2
using cubic spline first. The black bands in the figures overlaid
over the data points are such interpolations.  By choosing the
endpoint of the integration of the interpolated data to be either
1 or 2, we can get the free energy to introduce the $Q=1$
monopole-antimonopole pair, or the $Q=2$ monopole-antimonopole pair
respectively.  Thus, without an extra computational cost, we study
both $Q=1$ and $Q=2$ monopoles in this paper.

The numerical integration of the data results in the lattice free
energy, $F_\lat(\ell;L)$. In \fgn{fbare}, we show the $\ell$
dependence of $F_\lat(\ell;L)$ for $Q=1,2$ and $N=4,12$. The different
colored symbols within the panels are the data from different $L$.
At first sight,  the apparent decrease in $F_\lat(\ell;L)$ with an
increase in $\ell$ at various fixed $L$ might strike one to be
against expectation.  The reason behind such a behavior of the
lattice free energy is because the lattice spacing $\ell/L$ at
various $\ell$ at a fixed $L$ also changes when $\ell$ is increased.
The conversion of the lattice free energy to physical units should
restore a physically meaningful increasing tendency of the free
energy with the monopole-antimonopole separation, and also be able
to bring an approximate data collapse of the free energy from
different $L$.

We converted lattice correlator $G_\lat$ to physical $G_\phys$ by
a lattice spacing dependent factor $a^{-2D_Q^{(N)}}$ as explained
in \eqn{renormalization}.  Equivalently, the conversion between the
lattice and physical free energies is brought about by an additive
$2D_Q^{(N)} \log(a)$ term.  For regular composite operators built
out of the field operators $\psi$ and $A_\mu$ such as a fermion
bilinear $\bar\psi_i\bar\psi$, the ultraviolet dimensions follow
from the power-counting arguments; taking the example of fermion
bilinear, they are of ultraviolet dimension of two, and the lattice
bilinear can be converted to physical units by a factor $a^{-2}$.
However, a monopole operator at $x$ is not expressible in such a
simple form in terms of the fermion and gauge fields at $x$, and
power-counting cannot be performed. Therefore, we have to rely on
the empirical determination of the UV exponent $2D_Q^{(N)}$. As the
exponent $D_Q^{(N)}$ should govern the short-distance behavior of
the monopole-antimonopole correlator, we estimated $D_Q^{(N)}$ from
a leading logarithmic behavior,
\beq
F_\lat(L) = F_0 + 2 D_Q^{(N)} \log(L), 
\eeq{uvdimscaling}
of the lattice free energy at a fixed small lattice spacing $a=1/7$
corresponding to small box-sizes $a L \le 4$ on $L=12$ to 28. This
is equivalent to short monopole-antimonopole separations $t = aL/4
\le 1$ on such boxes where the above $\log(L)$ dependence could
arise.  In \fgn{uvdim}, we show such a $\log(L)$ dependence of
$F_\lat(L)$ at $a=1/7$ for $Q=1,2$ and $N=4,12$. The red data points
are from the Monte Carlo simulations on $L=$ 12, 16, 20, 24 and 28
lattices.  For $L\in[16,28]$, the data is consistent with a $\log(L)$
dependence of the free energy.  The $L=12$ lattice point is slightly
off from the logarithmic behavior, which suggests the presence of
lattice artifacts at such close $t=3a$ separation between the
monopole and the antimonopole. The black band is the best fit of
\eqn{uvdimscaling} to the data using $F_0$ and $D_Q^{(N)}$ as fit
parameters. Our best empirical estimates of the ultraviolet dimensions
are $D_1^{(4)}=1.85(11)$, $D_1^{(12)}=5.50(25)$, $D_2^{(4)}=4.67(16)$
and $D_2^{(12)}=14.01(36)$ respectively.

\bef
\centering
\includegraphics[scale=0.55]{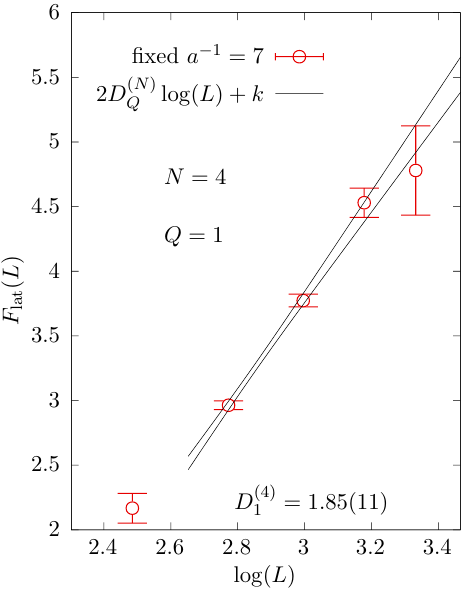}
\includegraphics[scale=0.55]{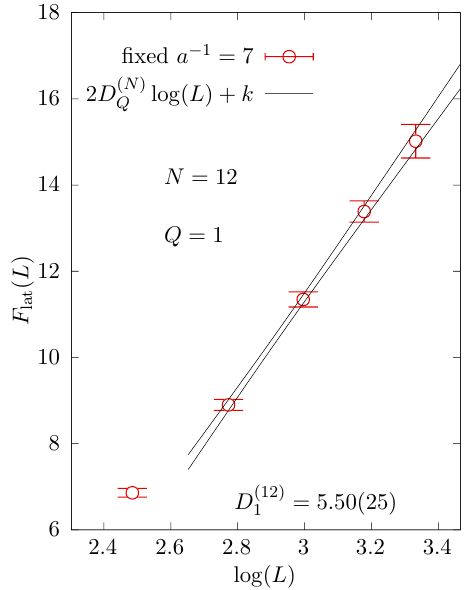}
\includegraphics[scale=0.55]{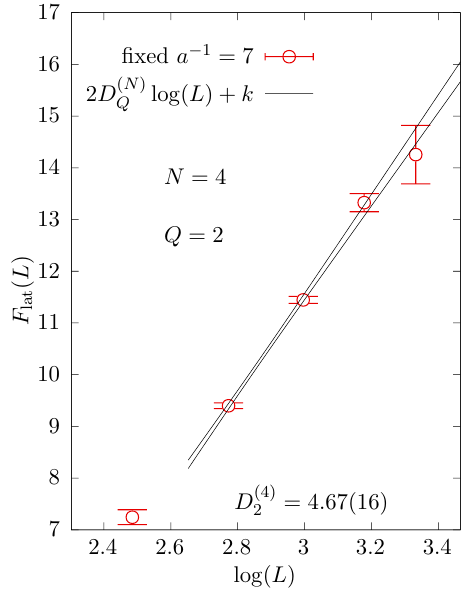}
\includegraphics[scale=0.55]{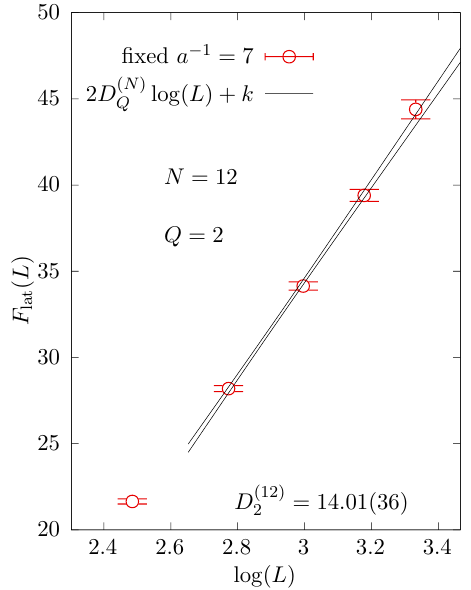}

\caption{ Estimation of UV scaling dimension $D^{(N)}_Q$ of the
flux-$Q$ monopoles from the scaling of the lattice free energy $F_\lat(L)$
in $N$-flavor theory with box size $L$ at fixed small lattice spacing
$\ell/L=1/7$.  The panels show $F_\lat(L)$ as a function of $L$.
The values of $(Q,N)$ in the panels from left to right correspond
to $(1,4), (1,12), (2,4), (2,12)$ respectively.  The data points
are the lattice-determined values of $F_\lat$.  The black bands are
the $\log(L)$ fits to the data. }
\eef{uvdim}

\bef
\centering
\includegraphics[scale=0.55]{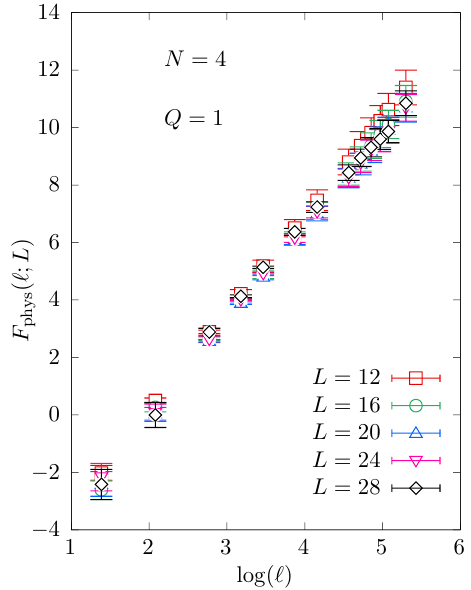}
\includegraphics[scale=0.55]{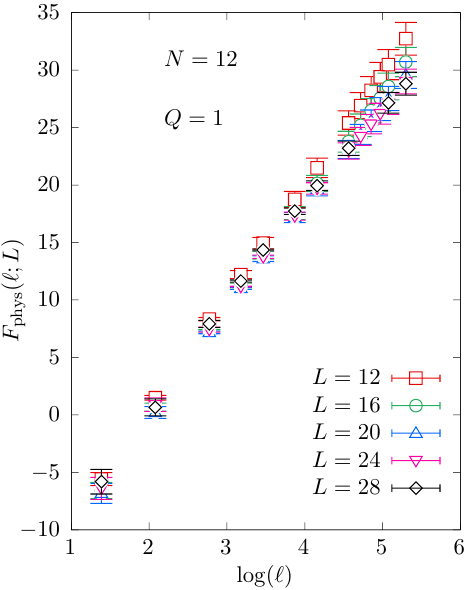}
\includegraphics[scale=0.55]{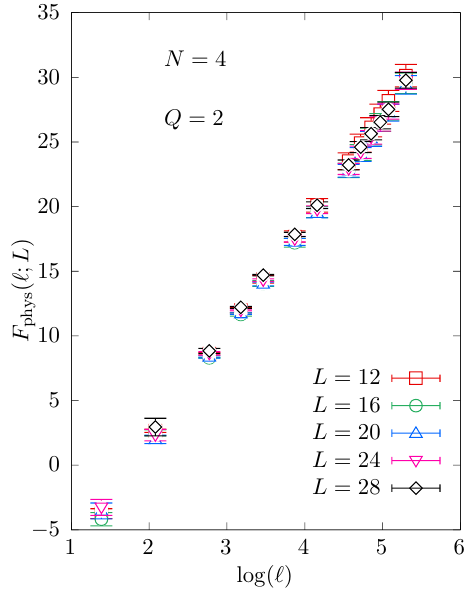}
\includegraphics[scale=0.55]{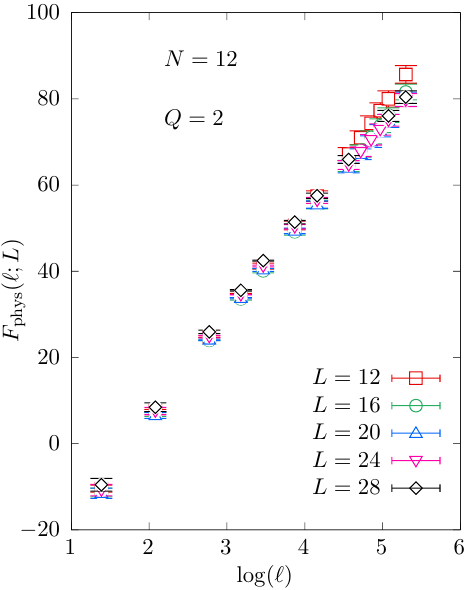}

\caption{The free energy $F_\phys(\ell,L)$ converted to physical
units is shown as a function of $\ell$. The data points at fixed
$L$ are shown using the different colored symbols. The values of
$(Q,N)$ in the panels from left to right correspond to $(1,4),
(1,12), (2,4), (2,12)$ respectively.}
\eef{fren}

\bef
\centering
\includegraphics[scale=0.55]{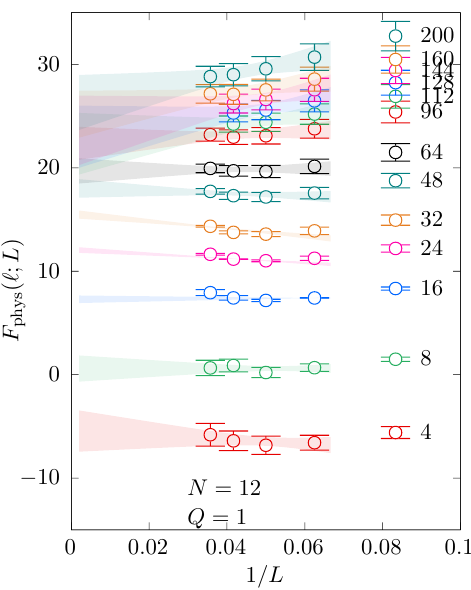}
\includegraphics[scale=0.55]{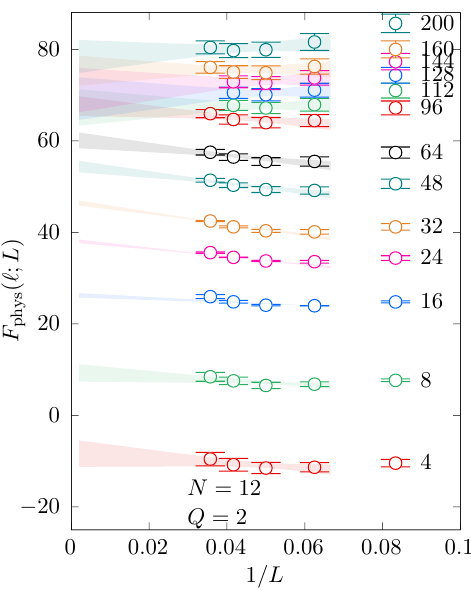}
\includegraphics[scale=0.55]{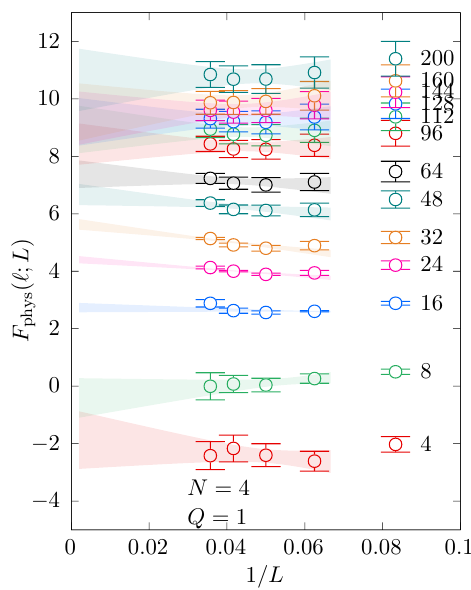}
\includegraphics[scale=0.55]{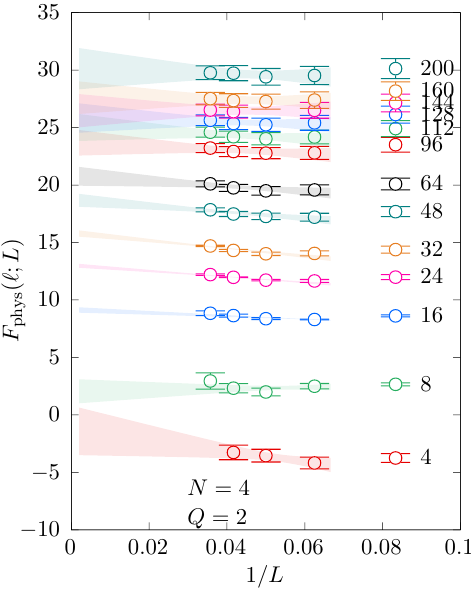}

\caption{The continuum estimates of the $F_\phys(\ell)$ through
$1/L$ extrapolations of $F_\phys(\ell;L)$ at different fixed physical
box sizes $\ell$ noted beside the data points. The $1/L$ fits were
made over the range $L\in[16,28]$. The bands show the extrapolations
resulting from the fits. The values of $(Q,N)$ for the panels from
left to right correspond to $(1,12), (2,12), (1,4), (2,4)$
respectively.}
\eef{cont}

Using the determined $F_\lat(\ell;L)$ and the best fit values of
$D_Q^{(N)}$ in the previous analysis, we obtained $F_\phys(\ell;L)$
as
\beq
F_\phys(\ell;L) = F_\lat(\ell;L) + 2D_Q^{(N)} \log\left(\frac{\ell}{L}\right).
\eeq{freelattoren}
We propagated the statistical errors in $F_\lat(\ell;L)$ and the
estimated $D_Q^{(N)}$ into the determination of $F_\phys$ by adding
the errors in quadrature.  In \fgn{fren}, we show $F_\phys(\ell;L)$
as a function of $\log(\ell)$ for $Q=1,2$ in $N=4$ and 12 flavor
theories. In each panel, we show the results of $F_\phys(\ell;L)$
from $L=12$, 16, 20, 24 and 28 together. First, we notice that
$F_\phys(\ell)$ increases monotonically as we expected.  Second,
the lattice-to-physical units `renormalization factor', $a^{-2D_Q^{(N)}}$
has caused a near data collapse of the $F_\phys(\ell;L)$ from
multiple $L$. The residual $L$ dependencies at fixed $\ell$ need
to be removed by extrapolating to $L\to\infty$ as we discuss below.

In \fgn{cont}, we show the residual $1/L$ dependence of $F_\phys(\ell;L)$
for $Q=1$ and 2 monopoles in $N=4$ and 12 theories.  The data points
differentiated by their colors have a fixed value of $\ell$, and
they have to be extrapolated to $L\to\infty$ to estimate the continuum
limit $F_\phys(\ell)$ in that physical box size. We perform the
extrapolation using a simple Ansatz, $F_\phys(\ell;L) = F_\phys(\ell)
+ k(\ell)/L$ with $F_\phys(\ell)$ and $k(\ell)$ fit parameters, to
describe the $L$-dependence of $F_\phys(\ell;L)$ for $L\in[16,28]$.
Such a fit was capable of describing the $L$-dependence well with
$\chi^2/{\rm dof}<1$ in most cases. For $\ell=112,128,144$ in $N=12$
theory,  we accidentally did not produce the $L=28$ lattice data.
Therefore, we performed the extrapolation only using $L=16,20$ and
24 data sets in those specific cases resulting in a comparatively
larger statistical error in their extrapolated values.  The various
colored bands in \fgn{cont} show the $1/L$ extrapolations at various
fixed $\ell$.  We will use the extrapolated $F_\phys(\ell)$ in the
discussion of infrared dimensions of monopole operators in the next
subsection.

\subsection{Estimation of infrared scaling dimensions}

First, we discuss the scaling dimensions in $N=12$ theory.  Due to
the relatively large value of $N$, this serves as a test case to
see if the values obtained for the scaling dimension agree approximately
with the large-$N$ expectations.  In \fgn{nf6q12cfren}, we show the
dependence of $F_\phys(\ell)$ as a function of $\log(\ell)$ for the
$N=12$ case.  In the left and right panels, we show the $Q=1$ and
$Q=2$ monopole free energies respectively.  The black points in the
panels are our estimates for the continuum limits of $F_\phys(\ell)$,
as obtained in \fgn{cont}. For comparison, we also show the data
points for $F_\phys(\ell)$ from $L=24$ lattice before performing
any continuum extrapolation.  One expects a simple $\log(\ell)$
dependence only in the large-box limit, corresponding to large
separations between the monopole and the antimonopole. Within the
statistical errors, we see such a $\log(\ell)$ dependence for
$\ell\ge 32$. We fitted
\beq
F_\phys(\ell) = f_0 + 2\Delta_Q^{(N)}\log(\ell),
\eeq{asymfitform}
using a constant $f_0$, and the infrared scaling dimension
$\Delta_Q^{(N)}$ as fit parameters over two ranges $\ell\in[32,200]$
and $\ell\in[48,200]$ to check for systematic dependence on fit
range.  The underlying lattice data for $F_\lat$ are statistically
independent at different $\ell$, but \eqn{freelattoren} introduces
correlations between different $\ell$ due to the commonality of the
second term in \eqn{freelattoren}. We found the covariance matrix
of the data for $F_\phys$ at different $\ell$ close to being singular
making the minimization of correlated $\chi^2$ to be not practical,
and we resorted to uncorrelated $\chi^2$ fits; this is an approximation
made in this study. We determined the statistical errors in fit
parameters using the Jack-knife method.  For the $N=12$ theory under
consideration, we determined $\Delta_1^{(12)}$ and $\Delta_2^{(12)}$
in this way. We show the resultant fits over $\ell\in[32,200]$ and
$[48,200]$ as the blue and magenta error bands respectively in the
two panels of \fgn{nf6q12cfren}.  The slopes of the $\log(\ell)$
behavior from the fits over $\ell\in[48,200]$ give
\beq
\Delta_1^{(12)}=2.81(66)\quad\text{and}\quad \Delta_2^{(12)}=8.2(1.0),
\eeq{n12q12deltavals}
for $Q=1$ and $Q=2$ monopoles respectively. The $\chi^2/{\rm dof}$
for the two fits are 1.1/6 and 3.2/6 respectively, which are smaller
than the typical value of around 1 due to the uncorrelated nature
of the fit.  By using a wider range of $\ell$ starting from a smaller
$\ell=32$, we found $\Delta_1^{(12)}=2.91(41)$ and
$\Delta_2^{(12)}=8.91(54)$ showing only a mild dependence on the
fit range.  The large-$N$ expectations~\cite{Pufu:2013vpa,Dyer:2013fja}
for these two scaling dimensions are $\Delta_1^{(12)}=3.1417$ and
$\Delta_2^{(12)}=7.882$.  We see that the estimates from the fit
performed over $\ell\in[48,200]$ to be quite consistent with the
large-$N$ expectation well within 1-$\sigma$ error.  The more precise
estimate of $\Delta_2^{(12)}$ from the fit over $\ell\in[32,200]$
is slightly higher than the large-$N$ value at the level of 2-$\sigma$,
and is more probable to be a systematic effect from using the smaller
$\ell=32$ in the fit rather than arise due to genuine higher $1/N$
corrections.  It is reassuring that our numerical method obtains
values for $\Delta_Q^{(12)}$ that are consistent with the large-$N$
expectations, which will add credence to the results from the method
at smaller $N$ to be discussed next.  As a minor note, by comparing
to the slope of the red points, we see that continuum extrapolation
at all $\ell$ was essential, without which we would have overestimated
the values of $\Delta_Q^{(12)}$ by instead fitting the $\ell$
dependence at a fixed $L$.

\bef
\centering
\includegraphics[scale=0.75]{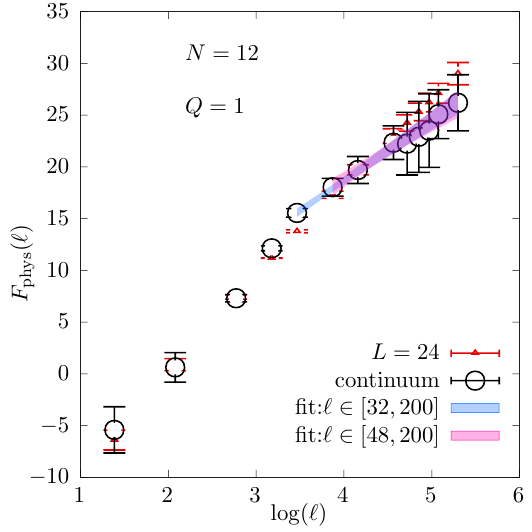}
\hskip 3em
\includegraphics[scale=0.75]{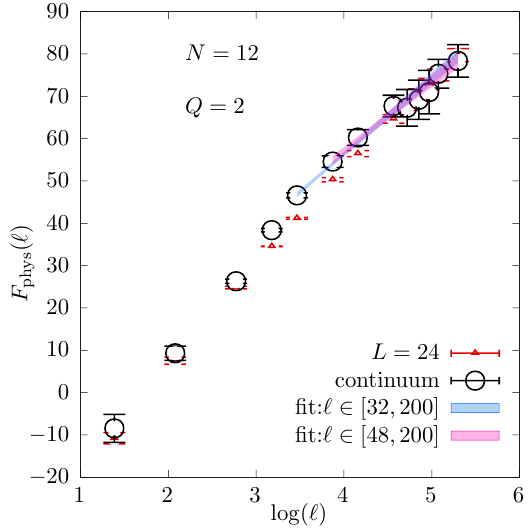}

\caption{ The physical free energy $F_\phys(\ell)$ in $N=12$ QED$_3$
is shown as a function of $\log(\ell)$ for $Q=1$ monopole in the
left panel and $Q=2$ monopole in the right panel.  The black points
are the continuum estimates and the red points are from a fixed
$L=24$. The blue and magenta bands are the $\log(\ell)$ fit over
$\ell\in[32,200]$ and $[48,200]$ respectively.}

\eef{nf6q12cfren}

\bef
\centering
\includegraphics[scale=0.75]{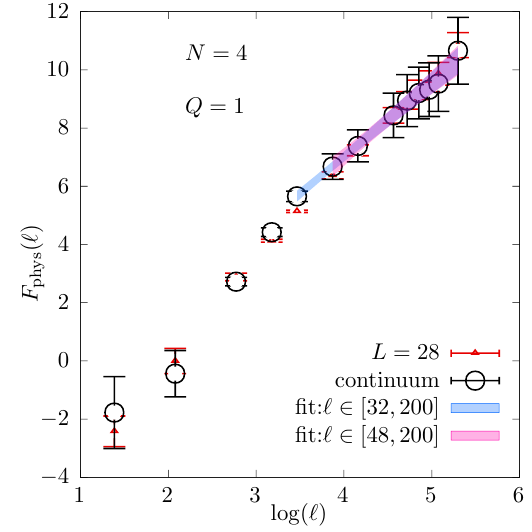}
\hskip 3em
\includegraphics[scale=0.75]{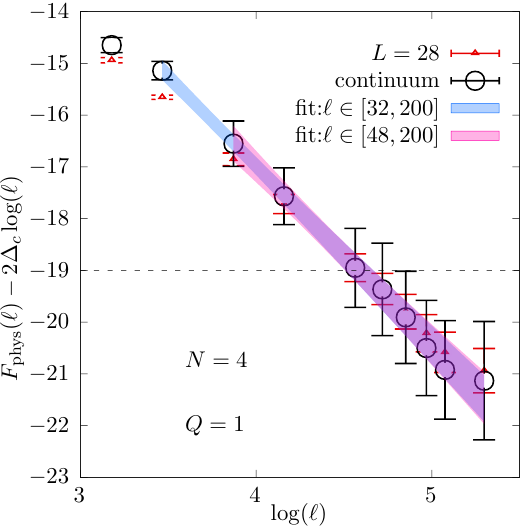}

\caption{ (Left panel) The free energy $F_\phys$ is shown as a
function of $\log(\ell)$ for $Q=1$ monopole in $N=4$ theory.  The
black points are the continuum estimates and the red points are
from fixed $L=28$ lattice.  The blue and magenta bands are the
$\log(\ell)$ fit over $\ell\in[32,200]$ and $[48,200]$ respectively.
The slopes of the fits give the scaling dimension $\Delta_1^{(4)}$.
(Right panel) The difference from the expectation of the marginally
relevant scaling with $\Delta_c=3$ is explored. The data points and
the fitted bands in the left panel at larger $\ell$ are re-plotted
as the difference, $F_\phys(\ell)-2\Delta_c\log(\ell)$. The horizontal
dashed line is shown to compare the residual slope in the data.  }

\eef{nf2q1cfren}

First, we consider the $Q=1$ monopole in the $N=4$ theory. We looked
at this case in our earlier work~\cite{Karthik:2019mrr}; the variation
in the present study is the usage of higher statistics, differences
in the sampled values of $\zeta$ to cover up to $\zeta=2$, and the
incorporation of dedicated continuum limits at each fixed $\ell$
instead of using a simpler one-parameter characterization of $1/L$
effects at all $\ell$ used in the earlier work.  In the left panel
of \fgn{nf2q1cfren}, we show the $\log(\ell)$ dependence of the
free energy for $Q=1$ monopole in the $N=4$ theory. The black points
are the continuum expectations, whereas the red ones are the data
from the largest $L=28$ lattice. Again, we see a simple $\log(\ell)$
behavior is consistent with the data from boxes with $\ell\in[32,200]$.
The fit to the functional form \eqn{asymfitform} over a range
$\ell\in[48,200]$ gives a slope of
\beq
\Delta_1^{(4)}=1.28(26),
\eeq{n4q1deltavals}
with a $\chi^2/{\rm dof}=1.2/6$.  We show the resulting fit as the
magenta error-band in \fgn{nf2q1cfren}. This is consistent with the
estimate $\Delta_1^{(4)}=1.25(9)$ from our earlier work. When we
include the smaller $\ell=32$ in the fit (shown as a blue band),
we find $\Delta_1^{(4)}=1.27(13)$ pointing to a very mild dependence
on fit range. Clearly, $\Delta_1^{(4)}$ is smaller than the marginal
value $\Delta_c=3$, which makes the $Q=1$ monopole operator relevant
along the renormalization group flows of $N=4$ QED$_3$. We can see
the relevance of $Q=1$ monopole operator without any fits by plotting
the difference $F_\phys(\ell)-2\Delta_c\log(\ell)$.  If the operator
is relevant, we should see a negative slope in the above difference.
Through a simple re-plotting of the data and fits in the left panel,
we show the $\log(\ell)$ dependence of the difference,
$F_\phys(\ell)-2\Delta_c\log(\ell)$, in the right panel. We see a
clear negative slope in the data and reach the same conclusion about
the relevance of $Q=1$ monopole in $N=4$ QED$_3$.  This brings us
to the main motivation for the present work; is the $Q=2$ monopole
operator also relevant in $N=4$ QED$_3$?

\bef
\centering
\includegraphics[scale=0.75]{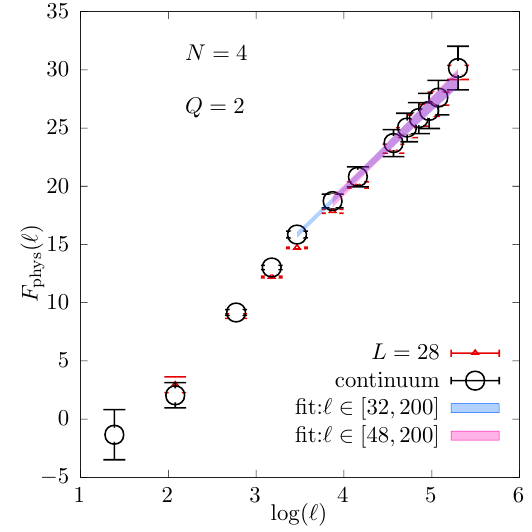}
\hskip 3em
\includegraphics[scale=0.75]{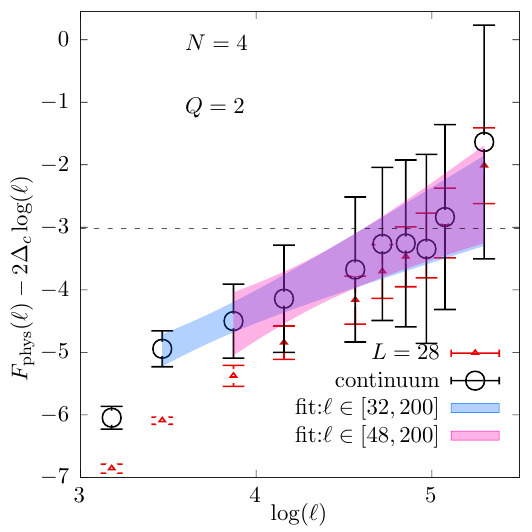}

\caption{ (Left panel) The free energy $F_\phys$ is shown as a
function of $\log(\ell)$ for $Q=2$ monopole in $N=4$ theory.  The
black points are the continuum estimates and the red points are
from fixed $L=28$ lattice.  The blue and magenta bands are the
$\log(\ell)$ fit over $\ell\in[32,200]$ and $[48,200]$ respectively.
The slopes of the fits give the scaling dimension $\Delta_2^{(4)}$.
(Right panel) The difference from the expectation of the marginally
relevant scaling with $\Delta_c=3$ is explored. The data points and
the fitted bands in the left panel at larger $\ell$ are re-plotted
as the difference, $F_\phys(\ell)-2\Delta_c\log(\ell)$. The horizontal
dashed line is shown to compare the residual slope in the data.
}
\eef{nf2q2cfren}

\bef
\centering
\includegraphics[scale=0.9]{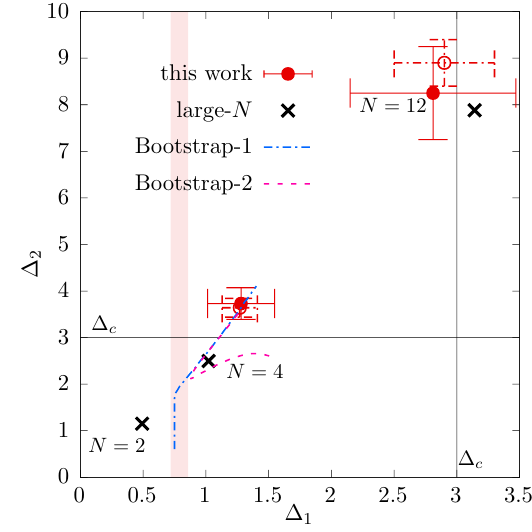}

\caption{ The scaling dimension of $Q=2$ monopole is plotted as a
function of $Q=1$ monopole using their estimates for $N=2,4$ and
12 QED$_3$.  The black crosses are the large-$N$ expectations.  The
filled red circles are the estimates in this paper obtained from
finite-size scaling analysis of the data over $\ell\in[48,200]$.
The open red circles are obtained by fitting the data over
$\ell\in[32,200]$.  The red vertical band is using the estimate of
$\Delta^{(2)}_1$ from our earlier work.  The two solid black lines
indicate the critical value of $\Delta_c=3$ for the two scaling
dimensions.  The blue dot-dashed line (taken
from~\cite{Chester:2016wrc,Chester:2017vdh} labelled Bootstrap-1)
and the magenta dashed line (taken from~\cite{Albayrak:2021xtd}
labelled Bootstrap-2) near the $N=4$ result are the boundaries
separating the allowed and disallowed regions using conformal
bootstrap methods for CFTs with infrared symmetry of $N=4$ QED$_3$.
The region below the blue dot-dashed line and the region enclosed
by the magenta line are the allowed regions according to the two
studies respectively, each imposing certain choices of constraints
on dimensions of operators occurring in the conformal expansion of
monopole four-point correlators. The endpoints of the boundaries
are simply the range shown in the two studies and not a hard cut-off.
}

\eef{delta}

In the left panel of \fgn{nf2q2cfren}, we show the $\log(\ell)$
dependence of $F_\phys(\ell)$ for $Q=2$ monopole in $N=4$ QED$_3$.
As in the previous cases we discussed, given the statistical errors,
the finite-size dependence of $F_\phys(\ell)$ for $\ell\ge 32$ is
consistent with a simple $\log(\ell)$ behavior. The magenta band
shows the fit using such a $\log(\ell)$ fit in \eqn{asymfitform}
to the data with $\ell\ge48$.  The value of the slope again gives
the scaling dimension. From the best fit values, we estimate the
scaling dimension of $Q=2$ monopole in $N=4$ QED$_3$ to be
\beq
\Delta_2^{(4)}=3.73(34),
\eeq{n4q2deltavals}
with $\chi^2/{\rm dof}=1.4/6$.  Thus, $\Delta_2^{(4)}>3$  with a
weak statistical significance of about 2-$\sigma$.  If we start the
fit from a smaller $\ell=32$, we find a similar value
$\Delta_2^{(4)}=3.65(21)$ with a smaller error.  At $O(1/N)$ in the
large-$N$ expansion, $\Delta_2^{(4)} = 2.498$. Our data allows the
possibility that either by the importance of higher $1/N$ orders
in the large-$N$ expansion or by a breakdown of such an expansion
for $N=4$, the value of $\Delta_2^{(4)}$ could be larger than 3,
and make it irrelevant in the infrared.  As we explained in the
previous case of $Q=1$ monopole, to argue that the data is consistent
with $\Delta_2^{(4)}>3$ without performing any fits, we re-plot the
data as a difference $F_\phys(\ell) - 2\Delta_c\log(\ell)$, where
the second term corresponds to the expected slope at a marginal
dimension $\Delta_c=3$.  In the right panel of \fgn{nf2q2cfren},
we show this difference over a range of larger $\ell$.  In this
plot, if the $Q=2$ monopole was relevant, one should see a $\log(\ell)$
dependence with a negative slope. The trend in the data indicates
a positive slope, which again points to the consistency of our data
with $Q=2$ monopole being irrelevant.  As a final remark, we note
that the behavior of $F_\phys(\ell)$ with $\Delta_2^{(4)}>3$ is not
strongly dependent on the continuum extrapolation procedure.  The
red points in the two panels of \fgn{nf2q2cfren} are the free
energies at different $\ell$ on the largest $L=28$ lattice.  From
the slope of the red points, we see that we would have reached an
even stronger conclusion that $\Delta_2^{(4)}>3$ from that data
alone. Therefore, the effect of $L\to\infty$ extrapolation has been
to make that conclusion weaker.

We collect the results of $\Delta_Q^{(N)}$ in \fgn{delta}. We plot
$\Delta_2^{(N)}$ as a function of $\Delta_1^{(N)}$, making the
dependence on $N$ implicit.  The solid red points in \fgn{delta}
are the values determined in this paper using fits over data from
$\ell\in[48,200]$.  To show the systematic artifacts in the estimate,
we also show the estimates from fits over data from $\ell\in[32,200]$
as the open circles.  In the previous work~\cite{Karthik:2019mrr},
we determined only the value of $\Delta_1^{(2)}$ in  $N=2$ QED$_3$.
Therefore, we show a red band in \fgn{delta} to indicate lack of
data for $\Delta_2^{(2)}$.  We show the large-$N$ expectation for
$\Delta_2^{(N)}$ versus $\Delta_1^{(N)}$ as the black crosses.  As
we discussed before, the top-right red point from $N=12$ QED$_3$
is consistent with the large-$N$ expectation.  The vertical and
horizontal dashed lines in the figure indicate the marginal values
of $\Delta_1=3$ and $\Delta_2=3$ respectively.  The data point from
$N=12$ lies at the edge of $\Delta_1=3$, which indicates that $N=12$
QED$_3$ is close to being the critical flavor below which $Q=1$
monopole becomes irrelevant. As pointed out in~\cite{Pufu:2013vpa},
one could conjecture that the critical flavor that separates the
mass-gapped and conformal infrared phases of $N$-flavor compact
QED$_3$, where all flux-$Q$ monopoles can freely arise, is around
$N\approx 12$.  The important finding in this paper is that the
$N=4$ data point in \fgn{delta} lies above the critical horizontal
line, albeit with a weaker statistical significance of about
2-$\sigma$ (or 3-$\sigma$ if one bases the conclusion on the red
open point).  We also see that the estimated location of the $N=4$
data point in the plot is quite robust with respect to change to
the fitted range of $\ell$.  Thus, our data cannot rule out the
scenario where $Q=2$ monopole remains irrelevant along the
renormalization group flow even for the $N=4$ theory.  For comparison,
we also show the boundary of the allowed region in the $\Delta_1-\Delta_2$
plane for CFTs that have the symmetries of $N=4$ QED$_3$ as determined
using the conformal bootstrap approach. The blue dot-dashed line
is the boundary computed in Refs~\cite{Chester:2016wrc,Chester:2017vdh},
and the region below the line is allowed. The magenta dashed line
is the boundary computed in Ref~\cite{Albayrak:2021xtd} with the
allowed region enclosed by the curve. In both cases, we obtained
the data from the plots as shown in the two papers and the region
covered in the $\Delta_1-\Delta_2$ plane is only representative of
the region shown in the two studies and not a hard cut-off~\footnote{
Furthermore, the allowed region is dependent on constraints imposed
on dimensions of operators that occur in the conformal expansion
of four-point functions used in the conformal bootstrap; we refer
the reader to Refs~\cite{Chester:2016wrc,Chester:2017vdh} and
Ref~\cite{Albayrak:2021xtd} for details of various constraints
imposed in the two studies. The blue line in \fgn{delta} corresponds
to the case with $\Delta_2\ge 2$ for $R = T, l = 0$ operator shown
in upper panel of Figure~7 of~\cite{Chester:2017vdh}. The purple
line in \fgn{delta} corresponds to the $\Delta_{\bar{AA}}\ge 2.4$
shown in Figure~10 of~\cite{Albayrak:2021xtd}.  The meaning of the
constraints and the nomenclature are discussed in the cited papers.
}.  In both the bootstrap studies, there exists an allowed region
where $\Delta_2>3$.  The data point for $N=4$ case from our lattice
study is quite consistent with this allowed region within errors,
and the central value conspicuously sits right at the allowed upper
boundary line in the two studies.  It would be interesting to fold
in this finding as an input for future conformal bootstrap studies.

\section{Conclusions}

Along with the composite operators such as fermion bilinears and
four-Fermi operators, monopole operators that introduce $2\pi Q$
fluxes around their insertion point constitute nontrivial insertions
in QED$_3$. The motivation for this study was the question of
infrared relevance of the $Q=2$ monopole operators in QED$_3$ coupled
to massless $N=4$ Dirac fermion flavors.  We used numerical lattice
simulations of noncompact QED$_3$ coupled to $N=4$ and $N=12$ flavors
of Wilson-Dirac fermions fine-tuned to the massless point. We
estimated the infrared scaling dimensions of $Q=1$ and $Q=2$ monopoles
in the $N=4$ and 12 theories from the finite-size scaling analysis
of free energy required to introduce the $Q=1$ and 2 monopole-antimonopole
pairs in the two theories.  We validated the method in $N=12$ theory
first where the values of the $Q=1$ and 2 scaling dimensions would
be expected to lie closer to the values obtained from the first-order
large-$N$ expansion. Then, by applying to the $N=4$ theory, we found
our best estimate for $Q=2$ scaling dimension to $3.73(34)$, which
is consistent with being greater than the marginal value of $3$.
Thus, our result favors, and certainly cannot rule out, the possibility
of $Q=2$ monopole operators being irrelevant at the infrared fixed
point of $N=4$ QED$_3$.  We summarized our results for the scaling
dimensions in \fgn{delta} that shows the dimension of $Q=2$ monopole
as a function of the dimension of $Q=1$ monopole, and we compared it 
to determinations from conformal bootstrap.

As argued in Refs~\cite{Song:2018ial,Song:2018ccm}, the irrelevance
of $Q=2$ monopole operators at the infrared fixed point of $N=4$
noncompact QED$_3$ could imply the possibility of hosting a stable
U(1) Dirac spin liquid phase in non-bipartite lattices, such as on
the triangular and Kagom\'e lattice. On such lattices, it has been
argued~\cite{Song:2018ial,Song:2018ial} that the $Q=1$ monopoles
are disallowed due to symmetry reasons, and the most important
destabilizing perturbation could be that of the next allowed
higher-flux monopole, which is the $Q=2$ monopole on the Kagom\'e
lattice.  The findings from our numerical study mildly support the
exciting possibility that the higher-flux monopoles might not
destabilize the Dirac spin liquid on such non-bipartite lattices.

\acknowledgments
The authors thank Yin-Chen He and Chong Wang for useful discussions.
The authors also thank Shai Chester for the discussion on the
conformal bootstrap results for monopole dimensions.  R.N. acknowledges
partial support by the NSF under grant number PHY-1913010 and
PHY-2310479.  This work used Expanse at SDSC through allocation
PHY220077 from the Advanced Cyberinfrastructure Coordination
Ecosystem: Services \& Support (ACCESS) program, which is supported
by National Science Foundation grants \#2138259, \#2138286, \#2138307,
\#2137603, and \#2138296.

\bibliography{biblio}

\begin{thebibliography}{41}%
\makeatletter
\providecommand \@ifxundefined [1]{%
 \@ifx{#1\undefined}
}%
\providecommand \@ifnum [1]{%
 \ifnum #1\expandafter \@firstoftwo
 \else \expandafter \@secondoftwo
 \fi
}%
\providecommand \@ifx [1]{%
 \ifx #1\expandafter \@firstoftwo
 \else \expandafter \@secondoftwo
 \fi
}%
\providecommand \natexlab [1]{#1}%
\providecommand \enquote  [1]{``#1''}%
\providecommand \bibnamefont  [1]{#1}%
\providecommand \bibfnamefont [1]{#1}%
\providecommand \citenamefont [1]{#1}%
\providecommand \href@noop [0]{\@secondoftwo}%
\providecommand \href [0]{\begingroup \@sanitize@url \@href}%
\providecommand \@href[1]{\@@startlink{#1}\@@href}%
\providecommand \@@href[1]{\endgroup#1\@@endlink}%
\providecommand \@sanitize@url [0]{\catcode `\\12\catcode `\$12\catcode
  `\&12\catcode `\#12\catcode `\^12\catcode `\_12\catcode `\%12\relax}%
\providecommand \@@startlink[1]{}%
\providecommand \@@endlink[0]{}%
\providecommand \url  [0]{\begingroup\@sanitize@url \@url }%
\providecommand \@url [1]{\endgroup\@href {#1}{\urlprefix }}%
\providecommand \urlprefix  [0]{URL }%
\providecommand \Eprint [0]{\href }%
\providecommand \doibase [0]{http://dx.doi.org/}%
\providecommand \selectlanguage [0]{\@gobble}%
\providecommand \bibinfo  [0]{\@secondoftwo}%
\providecommand \bibfield  [0]{\@secondoftwo}%
\providecommand \translation [1]{[#1]}%
\providecommand \BibitemOpen [0]{}%
\providecommand \bibitemStop [0]{}%
\providecommand \bibitemNoStop [0]{.\EOS\space}%
\providecommand \EOS [0]{\spacefactor3000\relax}%
\providecommand \BibitemShut  [1]{\csname bibitem#1\endcsname}%
\let\auto@bib@innerbib\@empty
%</preamble>
\bibitem [{\citenamefont {Hands}\ and\ \citenamefont
  {Kogut}(1990)}]{Hands:1989mv}%
  \BibitemOpen
  \bibfield  {author} {\bibinfo {author} {\bibfnamefont {S.}~\bibnamefont
  {Hands}}\ and\ \bibinfo {author} {\bibfnamefont {J.~B.}\ \bibnamefont
  {Kogut}},\ }\href {\doibase 10.1016/0550-3213(90)90503-6} {\bibfield
  {journal} {\bibinfo  {journal} {Nucl. Phys.}\ }\textbf {\bibinfo {volume}
  {B335}},\ \bibinfo {pages} {455} (\bibinfo {year} {1990})}\BibitemShut
  {NoStop}%
%%CITATION = NUPHA,B335,455;%%
\bibitem [{\citenamefont {Hands}\ \emph {et~al.}(2002)\citenamefont {Hands},
  \citenamefont {Kogut},\ and\ \citenamefont {Strouthos}}]{Hands:2002dv}%
  \BibitemOpen
  \bibfield  {author} {\bibinfo {author} {\bibfnamefont {S.}~\bibnamefont
  {Hands}}, \bibinfo {author} {\bibfnamefont {J.}~\bibnamefont {Kogut}}, \ and\
  \bibinfo {author} {\bibfnamefont {C.}~\bibnamefont {Strouthos}},\ }\href
  {\doibase 10.1016/S0550-3213(02)00869-6} {\bibfield  {journal} {\bibinfo
  {journal} {Nucl.Phys.}\ }\textbf {\bibinfo {volume} {B645}},\ \bibinfo
  {pages} {321} (\bibinfo {year} {2002})},\ \Eprint
  {http://arxiv.org/abs/hep-lat/0208030} {arXiv:hep-lat/0208030 [hep-lat]}
  \BibitemShut {NoStop}%
%%CITATION = HEP-LAT/0208030;%%
\bibitem [{\citenamefont {Hands}\ \emph {et~al.}(2004)\citenamefont {Hands},
  \citenamefont {Kogut}, \citenamefont {Scorzato},\ and\ \citenamefont
  {Strouthos}}]{Hands:2004bh}%
  \BibitemOpen
  \bibfield  {author} {\bibinfo {author} {\bibfnamefont {S.}~\bibnamefont
  {Hands}}, \bibinfo {author} {\bibfnamefont {J.}~\bibnamefont {Kogut}},
  \bibinfo {author} {\bibfnamefont {L.}~\bibnamefont {Scorzato}}, \ and\
  \bibinfo {author} {\bibfnamefont {C.}~\bibnamefont {Strouthos}},\ }\href
  {\doibase 10.1103/PhysRevB.70.104501} {\bibfield  {journal} {\bibinfo
  {journal} {Phys.Rev.}\ }\textbf {\bibinfo {volume} {B70}},\ \bibinfo {pages}
  {104501} (\bibinfo {year} {2004})},\ \Eprint
  {http://arxiv.org/abs/hep-lat/0404013} {arXiv:hep-lat/0404013 [hep-lat]}
  \BibitemShut {NoStop}%
%%CITATION = HEP-LAT/0404013;%%
\bibitem [{\citenamefont {Raviv}\ \emph {et~al.}(2014)\citenamefont {Raviv},
  \citenamefont {Shamir},\ and\ \citenamefont {Svetitsky}}]{Raviv:2014xna}%
  \BibitemOpen
  \bibfield  {author} {\bibinfo {author} {\bibfnamefont {O.}~\bibnamefont
  {Raviv}}, \bibinfo {author} {\bibfnamefont {Y.}~\bibnamefont {Shamir}}, \
  and\ \bibinfo {author} {\bibfnamefont {B.}~\bibnamefont {Svetitsky}},\ }\href
  {\doibase 10.1103/PhysRevD.90.014512} {\bibfield  {journal} {\bibinfo
  {journal} {Phys. Rev.}\ }\textbf {\bibinfo {volume} {D90}},\ \bibinfo {pages}
  {014512} (\bibinfo {year} {2014})},\ \Eprint {http://arxiv.org/abs/1405.6916}
  {arXiv:1405.6916 [hep-lat]} \BibitemShut {NoStop}%
%%CITATION = ARXIV:1405.6916;%%
\bibitem [{\citenamefont {Chester}\ and\ \citenamefont
  {Pufu}(2016)}]{Chester:2016wrc}%
  \BibitemOpen
  \bibfield  {author} {\bibinfo {author} {\bibfnamefont {S.~M.}\ \bibnamefont
  {Chester}}\ and\ \bibinfo {author} {\bibfnamefont {S.~S.}\ \bibnamefont
  {Pufu}},\ }\href {\doibase 10.1007/JHEP08(2016)019} {\bibfield  {journal}
  {\bibinfo  {journal} {JHEP}\ }\textbf {\bibinfo {volume} {08}},\ \bibinfo
  {pages} {019} (\bibinfo {year} {2016})},\ \Eprint
  {http://arxiv.org/abs/1601.03476} {arXiv:1601.03476 [hep-th]} \BibitemShut
  {NoStop}%
%%CITATION = ARXIV:1601.03476;%%
\bibitem [{\citenamefont {Chester}\ \emph {et~al.}(2018)\citenamefont
  {Chester}, \citenamefont {Iliesiu}, \citenamefont {Mezei},\ and\
  \citenamefont {Pufu}}]{Chester:2017vdh}%
  \BibitemOpen
  \bibfield  {author} {\bibinfo {author} {\bibfnamefont {S.~M.}\ \bibnamefont
  {Chester}}, \bibinfo {author} {\bibfnamefont {L.~V.}\ \bibnamefont
  {Iliesiu}}, \bibinfo {author} {\bibfnamefont {M.}~\bibnamefont {Mezei}}, \
  and\ \bibinfo {author} {\bibfnamefont {S.~S.}\ \bibnamefont {Pufu}},\ }\href
  {\doibase 10.1007/JHEP05(2018)157} {\bibfield  {journal} {\bibinfo  {journal}
  {JHEP}\ }\textbf {\bibinfo {volume} {05}},\ \bibinfo {pages} {157} (\bibinfo
  {year} {2018})},\ \Eprint {http://arxiv.org/abs/1710.00654} {arXiv:1710.00654
  [hep-th]} \BibitemShut {NoStop}%
\bibitem [{\citenamefont {He}\ \emph {et~al.}(2022)\citenamefont {He},
  \citenamefont {Rong},\ and\ \citenamefont {Su}}]{He:2021sto}%
  \BibitemOpen
  \bibfield  {author} {\bibinfo {author} {\bibfnamefont {Y.-C.}\ \bibnamefont
  {He}}, \bibinfo {author} {\bibfnamefont {J.}~\bibnamefont {Rong}}, \ and\
  \bibinfo {author} {\bibfnamefont {N.}~\bibnamefont {Su}},\ }\href {\doibase
  10.21468/SciPostPhys.13.2.014} {\bibfield  {journal} {\bibinfo  {journal}
  {SciPost Phys.}\ }\textbf {\bibinfo {volume} {13}},\ \bibinfo {pages} {014}
  (\bibinfo {year} {2022})},\ \Eprint {http://arxiv.org/abs/2107.14637}
  {arXiv:2107.14637 [cond-mat.str-el]} \BibitemShut {NoStop}%
\bibitem [{\citenamefont {Li}(2022)}]{Li:2021emd}%
  \BibitemOpen
  \bibfield  {author} {\bibinfo {author} {\bibfnamefont {Z.}~\bibnamefont
  {Li}},\ }\href {\doibase 10.1016/j.physletb.2022.137192} {\bibfield
  {journal} {\bibinfo  {journal} {Phys. Lett. B}\ }\textbf {\bibinfo {volume}
  {831}},\ \bibinfo {pages} {137192} (\bibinfo {year} {2022})},\ \Eprint
  {http://arxiv.org/abs/2107.09020} {arXiv:2107.09020 [hep-th]} \BibitemShut
  {NoStop}%
\bibitem [{\citenamefont {Albayrak}\ \emph {et~al.}(2022)\citenamefont
  {Albayrak}, \citenamefont {Erramilli}, \citenamefont {Li}, \citenamefont
  {Poland},\ and\ \citenamefont {Xin}}]{Albayrak:2021xtd}%
  \BibitemOpen
  \bibfield  {author} {\bibinfo {author} {\bibfnamefont {S.}~\bibnamefont
  {Albayrak}}, \bibinfo {author} {\bibfnamefont {R.~S.}\ \bibnamefont
  {Erramilli}}, \bibinfo {author} {\bibfnamefont {Z.}~\bibnamefont {Li}},
  \bibinfo {author} {\bibfnamefont {D.}~\bibnamefont {Poland}}, \ and\ \bibinfo
  {author} {\bibfnamefont {Y.}~\bibnamefont {Xin}},\ }\href {\doibase
  10.1103/PhysRevD.105.085008} {\bibfield  {journal} {\bibinfo  {journal}
  {Phys. Rev. D}\ }\textbf {\bibinfo {volume} {105}},\ \bibinfo {pages}
  {085008} (\bibinfo {year} {2022})},\ \Eprint
  {http://arxiv.org/abs/2112.02106} {arXiv:2112.02106 [hep-th]} \BibitemShut
  {NoStop}%
\bibitem [{\citenamefont {Rychkov}\ and\ \citenamefont
  {Su}(2023)}]{Rychkov:2023wsd}%
  \BibitemOpen
  \bibfield  {author} {\bibinfo {author} {\bibfnamefont {S.}~\bibnamefont
  {Rychkov}}\ and\ \bibinfo {author} {\bibfnamefont {N.}~\bibnamefont {Su}},\
  }\href@noop {} {\  (\bibinfo {year} {2023})},\ \Eprint
  {http://arxiv.org/abs/2311.15844} {arXiv:2311.15844 [hep-th]} \BibitemShut
  {NoStop}%
\bibitem [{\citenamefont {Pisarski}(1984)}]{Pisarski:1984dj}%
  \BibitemOpen
  \bibfield  {author} {\bibinfo {author} {\bibfnamefont {R.~D.}\ \bibnamefont
  {Pisarski}},\ }\href {\doibase 10.1103/PhysRevD.29.2423} {\bibfield
  {journal} {\bibinfo  {journal} {Phys.Rev.}\ }\textbf {\bibinfo {volume}
  {D29}},\ \bibinfo {pages} {2423} (\bibinfo {year} {1984})}\BibitemShut
  {NoStop}%
%%CITATION = PHRVA,D29,2423;%%
\bibitem [{\citenamefont {Appelquist}\ \emph {et~al.}(1985)\citenamefont
  {Appelquist}, \citenamefont {Bowick}, \citenamefont {Cohler},\ and\
  \citenamefont {Wijewardhana}}]{Appelquist:1985vf}%
  \BibitemOpen
  \bibfield  {author} {\bibinfo {author} {\bibfnamefont {T.}~\bibnamefont
  {Appelquist}}, \bibinfo {author} {\bibfnamefont {M.~J.}\ \bibnamefont
  {Bowick}}, \bibinfo {author} {\bibfnamefont {E.}~\bibnamefont {Cohler}}, \
  and\ \bibinfo {author} {\bibfnamefont {L.~C.~R.}\ \bibnamefont
  {Wijewardhana}},\ }\href {\doibase 10.1103/PhysRevLett.55.1715} {\bibfield
  {journal} {\bibinfo  {journal} {Phys. Rev. Lett.}\ }\textbf {\bibinfo
  {volume} {55}},\ \bibinfo {pages} {1715} (\bibinfo {year}
  {1985})}\BibitemShut {NoStop}%
%%CITATION = PRLTA,55,1715;%%
\bibitem [{\citenamefont {Appelquist}\ \emph
  {et~al.}(1986{\natexlab{a}})\citenamefont {Appelquist}, \citenamefont
  {Bowick}, \citenamefont {Karabali},\ and\ \citenamefont
  {Wijewardhana}}]{Appelquist:1986qw}%
  \BibitemOpen
  \bibfield  {author} {\bibinfo {author} {\bibfnamefont {T.}~\bibnamefont
  {Appelquist}}, \bibinfo {author} {\bibfnamefont {M.~J.}\ \bibnamefont
  {Bowick}}, \bibinfo {author} {\bibfnamefont {D.}~\bibnamefont {Karabali}}, \
  and\ \bibinfo {author} {\bibfnamefont {L.~C.~R.}\ \bibnamefont
  {Wijewardhana}},\ }\href {\doibase 10.1103/PhysRevD.33.3774} {\bibfield
  {journal} {\bibinfo  {journal} {Phys. Rev.}\ }\textbf {\bibinfo {volume}
  {D33}},\ \bibinfo {pages} {3774} (\bibinfo {year}
  {1986}{\natexlab{a}})}\BibitemShut {NoStop}%
%%CITATION = PHRVA,D33,3774;%%
\bibitem [{\citenamefont {Appelquist}\ \emph
  {et~al.}(1986{\natexlab{b}})\citenamefont {Appelquist}, \citenamefont
  {Bowick}, \citenamefont {Karabali},\ and\ \citenamefont
  {Wijewardhana}}]{Appelquist:1986fd}%
  \BibitemOpen
  \bibfield  {author} {\bibinfo {author} {\bibfnamefont {T.~W.}\ \bibnamefont
  {Appelquist}}, \bibinfo {author} {\bibfnamefont {M.~J.}\ \bibnamefont
  {Bowick}}, \bibinfo {author} {\bibfnamefont {D.}~\bibnamefont {Karabali}}, \
  and\ \bibinfo {author} {\bibfnamefont {L.~C.~R.}\ \bibnamefont
  {Wijewardhana}},\ }\href {\doibase 10.1103/PhysRevD.33.3704} {\bibfield
  {journal} {\bibinfo  {journal} {Phys. Rev.}\ }\textbf {\bibinfo {volume}
  {D33}},\ \bibinfo {pages} {3704} (\bibinfo {year}
  {1986}{\natexlab{b}})}\BibitemShut {NoStop}%
%%CITATION = PHRVA,D33,3704;%%
\bibitem [{\citenamefont {Appelquist}\ \emph {et~al.}(1988)\citenamefont
  {Appelquist}, \citenamefont {Nash},\ and\ \citenamefont
  {Wijewardhana}}]{Appelquist:1988sr}%
  \BibitemOpen
  \bibfield  {author} {\bibinfo {author} {\bibfnamefont {T.}~\bibnamefont
  {Appelquist}}, \bibinfo {author} {\bibfnamefont {D.}~\bibnamefont {Nash}}, \
  and\ \bibinfo {author} {\bibfnamefont {L.~C.~R.}\ \bibnamefont
  {Wijewardhana}},\ }\href {\doibase 10.1103/PhysRevLett.60.2575} {\bibfield
  {journal} {\bibinfo  {journal} {Phys. Rev. Lett.}\ }\textbf {\bibinfo
  {volume} {60}},\ \bibinfo {pages} {2575} (\bibinfo {year}
  {1988})}\BibitemShut {NoStop}%
%%CITATION = PRLTA,60,2575;%%
\bibitem [{\citenamefont {Gusynin}\ and\ \citenamefont
  {Reenders}(2003)}]{Gusynin:2003ww}%
  \BibitemOpen
  \bibfield  {author} {\bibinfo {author} {\bibfnamefont {V.~P.}\ \bibnamefont
  {Gusynin}}\ and\ \bibinfo {author} {\bibfnamefont {M.}~\bibnamefont
  {Reenders}},\ }\href {\doibase 10.1103/PhysRevD.68.025017} {\bibfield
  {journal} {\bibinfo  {journal} {Phys. Rev.}\ }\textbf {\bibinfo {volume}
  {D68}},\ \bibinfo {pages} {025017} (\bibinfo {year} {2003})},\ \Eprint
  {http://arxiv.org/abs/hep-ph/0304302} {arXiv:hep-ph/0304302 [hep-ph]}
  \BibitemShut {NoStop}%
%%CITATION = HEP-PH/0304302;%%
\bibitem [{\citenamefont {Gusynin}\ and\ \citenamefont
  {Pyatkovskiy}(2016)}]{Gusynin:2016som}%
  \BibitemOpen
  \bibfield  {author} {\bibinfo {author} {\bibfnamefont {V.~P.}\ \bibnamefont
  {Gusynin}}\ and\ \bibinfo {author} {\bibfnamefont {P.~K.}\ \bibnamefont
  {Pyatkovskiy}},\ }\href {\doibase 10.1103/PhysRevD.94.125009} {\bibfield
  {journal} {\bibinfo  {journal} {Phys. Rev.}\ }\textbf {\bibinfo {volume}
  {D94}},\ \bibinfo {pages} {125009} (\bibinfo {year} {2016})},\ \Eprint
  {http://arxiv.org/abs/1607.08582} {arXiv:1607.08582 [hep-ph]} \BibitemShut
  {NoStop}%
%%CITATION = ARXIV:1607.08582;%%
\bibitem [{\citenamefont {Kotikov}\ \emph {et~al.}(2016)\citenamefont
  {Kotikov}, \citenamefont {Shilin},\ and\ \citenamefont
  {Teber}}]{Kotikov:2016wrb}%
  \BibitemOpen
  \bibfield  {author} {\bibinfo {author} {\bibfnamefont {A.~V.}\ \bibnamefont
  {Kotikov}}, \bibinfo {author} {\bibfnamefont {V.~I.}\ \bibnamefont {Shilin}},
  \ and\ \bibinfo {author} {\bibfnamefont {S.}~\bibnamefont {Teber}},\ }\href
  {\doibase 10.1103/PhysRevD.94.056009} {\bibfield  {journal} {\bibinfo
  {journal} {Phys. Rev.}\ }\textbf {\bibinfo {volume} {D94}},\ \bibinfo {pages}
  {056009} (\bibinfo {year} {2016})},\ \Eprint
  {http://arxiv.org/abs/1605.01911} {arXiv:1605.01911 [hep-th]} \BibitemShut
  {NoStop}%
%%CITATION = ARXIV:1605.01911;%%
\bibitem [{\citenamefont {Karthik}\ and\ \citenamefont
  {Narayanan}(2016{\natexlab{a}})}]{Karthik:2015sgq}%
  \BibitemOpen
  \bibfield  {author} {\bibinfo {author} {\bibfnamefont {N.}~\bibnamefont
  {Karthik}}\ and\ \bibinfo {author} {\bibfnamefont {R.}~\bibnamefont
  {Narayanan}},\ }\href {\doibase 10.1103/PhysRevD.93.045020} {\bibfield
  {journal} {\bibinfo  {journal} {Phys. Rev.}\ }\textbf {\bibinfo {volume}
  {D93}},\ \bibinfo {pages} {045020} (\bibinfo {year} {2016}{\natexlab{a}})},\
  \Eprint {http://arxiv.org/abs/1512.02993} {arXiv:1512.02993 [hep-lat]}
  \BibitemShut {NoStop}%
%%CITATION = ARXIV:1512.02993;%%
\bibitem [{\citenamefont {Karthik}\ and\ \citenamefont
  {Narayanan}(2016{\natexlab{b}})}]{Karthik:2016ppr}%
  \BibitemOpen
  \bibfield  {author} {\bibinfo {author} {\bibfnamefont {N.}~\bibnamefont
  {Karthik}}\ and\ \bibinfo {author} {\bibfnamefont {R.}~\bibnamefont
  {Narayanan}},\ }\href {\doibase 10.1103/PhysRevD.94.065026} {\bibfield
  {journal} {\bibinfo  {journal} {Phys. Rev.}\ }\textbf {\bibinfo {volume}
  {D94}},\ \bibinfo {pages} {065026} (\bibinfo {year} {2016}{\natexlab{b}})},\
  \Eprint {http://arxiv.org/abs/1606.04109} {arXiv:1606.04109 [hep-th]}
  \BibitemShut {NoStop}%
%%CITATION = ARXIV:1606.04109;%%
\bibitem [{\citenamefont {Karthik}\ and\ \citenamefont
  {Narayanan}(2020)}]{Karthik:2020shl}%
  \BibitemOpen
  \bibfield  {author} {\bibinfo {author} {\bibfnamefont {N.}~\bibnamefont
  {Karthik}}\ and\ \bibinfo {author} {\bibfnamefont {R.}~\bibnamefont
  {Narayanan}},\ }\href {\doibase 10.1103/PhysRevLett.125.261601} {\bibfield
  {journal} {\bibinfo  {journal} {Phys. Rev. Lett.}\ }\textbf {\bibinfo
  {volume} {125}},\ \bibinfo {pages} {261601} (\bibinfo {year} {2020})},\
  \Eprint {http://arxiv.org/abs/2009.01313} {arXiv:2009.01313 [hep-lat]}
  \BibitemShut {NoStop}%
\bibitem [{\citenamefont {Karthik}\ and\ \citenamefont
  {Narayanan}(2017)}]{Karthik:2017hol}%
  \BibitemOpen
  \bibfield  {author} {\bibinfo {author} {\bibfnamefont {N.}~\bibnamefont
  {Karthik}}\ and\ \bibinfo {author} {\bibfnamefont {R.}~\bibnamefont
  {Narayanan}},\ }\href {\doibase 10.1103/PhysRevD.96.054509} {\bibfield
  {journal} {\bibinfo  {journal} {Phys. Rev.}\ }\textbf {\bibinfo {volume}
  {D96}},\ \bibinfo {pages} {054509} (\bibinfo {year} {2017})},\ \Eprint
  {http://arxiv.org/abs/1705.11143} {arXiv:1705.11143 [hep-th]} \BibitemShut
  {NoStop}%
%%CITATION = ARXIV:1705.11143;%%
\bibitem [{\citenamefont {Polyakov}(1975)}]{Polyakov:1975rs}%
  \BibitemOpen
  \bibfield  {author} {\bibinfo {author} {\bibfnamefont {A.~M.}\ \bibnamefont
  {Polyakov}},\ }\href {\doibase 10.1016/0370-2693(75)90162-8} {\bibfield
  {journal} {\bibinfo  {journal} {Phys. Lett.}\ }\textbf {\bibinfo {volume}
  {B59}},\ \bibinfo {pages} {82} (\bibinfo {year} {1975})},\ \bibinfo {note}
  {[,334(1975)]}\BibitemShut {NoStop}%
%%CITATION = PHLTA,B59,82;%%
\bibitem [{\citenamefont {Polyakov}(1977)}]{Polyakov:1976fu}%
  \BibitemOpen
  \bibfield  {author} {\bibinfo {author} {\bibfnamefont {A.~M.}\ \bibnamefont
  {Polyakov}},\ }\href {\doibase 10.1016/0550-3213(77)90086-4} {\bibfield
  {journal} {\bibinfo  {journal} {Nucl. Phys.}\ }\textbf {\bibinfo {volume}
  {B120}},\ \bibinfo {pages} {429} (\bibinfo {year} {1977})}\BibitemShut
  {NoStop}%
%%CITATION = NUPHA,B120,429;%%
\bibitem [{\citenamefont {Borokhov}\ \emph {et~al.}(2002)\citenamefont
  {Borokhov}, \citenamefont {Kapustin},\ and\ \citenamefont
  {Wu}}]{Borokhov:2002ib}%
  \BibitemOpen
  \bibfield  {author} {\bibinfo {author} {\bibfnamefont {V.}~\bibnamefont
  {Borokhov}}, \bibinfo {author} {\bibfnamefont {A.}~\bibnamefont {Kapustin}},
  \ and\ \bibinfo {author} {\bibfnamefont {X.-k.}\ \bibnamefont {Wu}},\ }\href
  {\doibase 10.1088/1126-6708/2002/11/049} {\bibfield  {journal} {\bibinfo
  {journal} {JHEP}\ }\textbf {\bibinfo {volume} {11}},\ \bibinfo {pages} {049}
  (\bibinfo {year} {2002})},\ \Eprint {http://arxiv.org/abs/hep-th/0206054}
  {arXiv:hep-th/0206054 [hep-th]} \BibitemShut {NoStop}%
%%CITATION = HEP-TH/0206054;%%
\bibitem [{\citenamefont {Karthik}\ and\ \citenamefont
  {Narayanan}(2019{\natexlab{a}})}]{Karthik:2019jds}%
  \BibitemOpen
  \bibfield  {author} {\bibinfo {author} {\bibfnamefont {N.}~\bibnamefont
  {Karthik}}\ and\ \bibinfo {author} {\bibfnamefont {R.}~\bibnamefont
  {Narayanan}},\ }\href {\doibase 10.1103/PhysRevD.100.094501} {\bibfield
  {journal} {\bibinfo  {journal} {Phys. Rev. D}\ }\textbf {\bibinfo {volume}
  {100}},\ \bibinfo {pages} {094501} (\bibinfo {year} {2019}{\natexlab{a}})},\
  \Eprint {http://arxiv.org/abs/1908.05284} {arXiv:1908.05284 [hep-lat]}
  \BibitemShut {NoStop}%
\bibitem [{\citenamefont {Hands}\ \emph {et~al.}(2006)\citenamefont {Hands},
  \citenamefont {Kogut},\ and\ \citenamefont {Lucini}}]{Hands:2006dh}%
  \BibitemOpen
  \bibfield  {author} {\bibinfo {author} {\bibfnamefont {S.}~\bibnamefont
  {Hands}}, \bibinfo {author} {\bibfnamefont {J.~B.}\ \bibnamefont {Kogut}}, \
  and\ \bibinfo {author} {\bibfnamefont {B.}~\bibnamefont {Lucini}},\
  }\href@noop {} {\  (\bibinfo {year} {2006})},\ \Eprint
  {http://arxiv.org/abs/hep-lat/0601001} {arXiv:hep-lat/0601001 [hep-lat]}
  \BibitemShut {NoStop}%
%%CITATION = HEP-LAT/0601001;%%
\bibitem [{\citenamefont {Armour}\ \emph {et~al.}(2011)\citenamefont {Armour},
  \citenamefont {Hands}, \citenamefont {Kogut}, \citenamefont {Lucini},
  \citenamefont {Strouthos},\ and\ \citenamefont {Vranas}}]{Armour:2011zx}%
  \BibitemOpen
  \bibfield  {author} {\bibinfo {author} {\bibfnamefont {W.}~\bibnamefont
  {Armour}}, \bibinfo {author} {\bibfnamefont {S.}~\bibnamefont {Hands}},
  \bibinfo {author} {\bibfnamefont {J.~B.}\ \bibnamefont {Kogut}}, \bibinfo
  {author} {\bibfnamefont {B.}~\bibnamefont {Lucini}}, \bibinfo {author}
  {\bibfnamefont {C.}~\bibnamefont {Strouthos}}, \ and\ \bibinfo {author}
  {\bibfnamefont {P.}~\bibnamefont {Vranas}},\ }\href {\doibase
  10.1103/PhysRevD.84.014502} {\bibfield  {journal} {\bibinfo  {journal} {Phys.
  Rev.}\ }\textbf {\bibinfo {volume} {D84}},\ \bibinfo {pages} {014502}
  (\bibinfo {year} {2011})},\ \Eprint {http://arxiv.org/abs/1105.3120}
  {arXiv:1105.3120 [hep-lat]} \BibitemShut {NoStop}%
%%CITATION = ARXIV:1105.3120;%%
\bibitem [{\citenamefont {Xu}\ \emph {et~al.}(2019)\citenamefont {Xu},
  \citenamefont {Qi}, \citenamefont {Zhang}, \citenamefont {Assaad},
  \citenamefont {Xu},\ and\ \citenamefont {Meng}}]{Xu:2018wyg}%
  \BibitemOpen
  \bibfield  {author} {\bibinfo {author} {\bibfnamefont {X.~Y.}\ \bibnamefont
  {Xu}}, \bibinfo {author} {\bibfnamefont {Y.}~\bibnamefont {Qi}}, \bibinfo
  {author} {\bibfnamefont {L.}~\bibnamefont {Zhang}}, \bibinfo {author}
  {\bibfnamefont {F.~F.}\ \bibnamefont {Assaad}}, \bibinfo {author}
  {\bibfnamefont {C.}~\bibnamefont {Xu}}, \ and\ \bibinfo {author}
  {\bibfnamefont {Z.~Y.}\ \bibnamefont {Meng}},\ }\href {\doibase
  10.1103/PhysRevX.9.021022} {\bibfield  {journal} {\bibinfo  {journal} {Phys.
  Rev.}\ }\textbf {\bibinfo {volume} {X9}},\ \bibinfo {pages} {021022}
  (\bibinfo {year} {2019})},\ \Eprint {http://arxiv.org/abs/1807.07574}
  {arXiv:1807.07574 [cond-mat.str-el]} \BibitemShut {NoStop}%
%%CITATION = ARXIV:1807.07574;%%
\bibitem [{\citenamefont {Pufu}(2014)}]{Pufu:2013vpa}%
  \BibitemOpen
  \bibfield  {author} {\bibinfo {author} {\bibfnamefont {S.~S.}\ \bibnamefont
  {Pufu}},\ }\href {\doibase 10.1103/PhysRevD.89.065016} {\bibfield  {journal}
  {\bibinfo  {journal} {Phys. Rev.}\ }\textbf {\bibinfo {volume} {D89}},\
  \bibinfo {pages} {065016} (\bibinfo {year} {2014})},\ \Eprint
  {http://arxiv.org/abs/1303.6125} {arXiv:1303.6125 [hep-th]} \BibitemShut
  {NoStop}%
%%CITATION = ARXIV:1303.6125;%%
\bibitem [{\citenamefont {Chester}\ \emph {et~al.}(2016)\citenamefont
  {Chester}, \citenamefont {Mezei}, \citenamefont {Pufu},\ and\ \citenamefont
  {Yaakov}}]{Chester:2015wao}%
  \BibitemOpen
  \bibfield  {author} {\bibinfo {author} {\bibfnamefont {S.~M.}\ \bibnamefont
  {Chester}}, \bibinfo {author} {\bibfnamefont {M.}~\bibnamefont {Mezei}},
  \bibinfo {author} {\bibfnamefont {S.~S.}\ \bibnamefont {Pufu}}, \ and\
  \bibinfo {author} {\bibfnamefont {I.}~\bibnamefont {Yaakov}},\ }\href
  {\doibase 10.1007/JHEP12(2016)015} {\bibfield  {journal} {\bibinfo  {journal}
  {JHEP}\ }\textbf {\bibinfo {volume} {12}},\ \bibinfo {pages} {015} (\bibinfo
  {year} {2016})},\ \Eprint {http://arxiv.org/abs/1511.07108} {arXiv:1511.07108
  [hep-th]} \BibitemShut {NoStop}%
%%CITATION = ARXIV:1511.07108;%%
\bibitem [{\citenamefont {Song}\ \emph {et~al.}(2020)\citenamefont {Song},
  \citenamefont {He}, \citenamefont {Vishwanath},\ and\ \citenamefont
  {Wang}}]{Song:2018ial}%
  \BibitemOpen
  \bibfield  {author} {\bibinfo {author} {\bibfnamefont {X.-Y.}\ \bibnamefont
  {Song}}, \bibinfo {author} {\bibfnamefont {Y.-C.}\ \bibnamefont {He}},
  \bibinfo {author} {\bibfnamefont {A.}~\bibnamefont {Vishwanath}}, \ and\
  \bibinfo {author} {\bibfnamefont {C.}~\bibnamefont {Wang}},\ }\href {\doibase
  10.1103/PhysRevX.10.011033} {\bibfield  {journal} {\bibinfo  {journal} {Phys.
  Rev. X}\ }\textbf {\bibinfo {volume} {10}},\ \bibinfo {pages} {011033}
  (\bibinfo {year} {2020})},\ \Eprint {http://arxiv.org/abs/1811.11182}
  {arXiv:1811.11182 [cond-mat.str-el]} \BibitemShut {NoStop}%
\bibitem [{\citenamefont {Song}\ \emph {et~al.}(2019)\citenamefont {Song},
  \citenamefont {Wang}, \citenamefont {Vishwanath},\ and\ \citenamefont
  {He}}]{Song:2018ccm}%
  \BibitemOpen
  \bibfield  {author} {\bibinfo {author} {\bibfnamefont {X.-Y.}\ \bibnamefont
  {Song}}, \bibinfo {author} {\bibfnamefont {C.}~\bibnamefont {Wang}}, \bibinfo
  {author} {\bibfnamefont {A.}~\bibnamefont {Vishwanath}}, \ and\ \bibinfo
  {author} {\bibfnamefont {Y.-C.}\ \bibnamefont {He}},\ }\href {\doibase
  10.1038/s41467-019-11727-3} {\bibfield  {journal} {\bibinfo  {journal}
  {Nature Commun.}\ }\textbf {\bibinfo {volume} {10}},\ \bibinfo {pages} {4254}
  (\bibinfo {year} {2019})},\ \Eprint {http://arxiv.org/abs/1811.11186}
  {arXiv:1811.11186 [cond-mat.str-el]} \BibitemShut {NoStop}%
\bibitem [{\citenamefont {Zhu}\ \emph {et~al.}(2018)\citenamefont {Zhu},
  \citenamefont {Chen}, \citenamefont {He},\ and\ \citenamefont
  {Witczak-Krempa}}]{Zhu:2018thc}%
  \BibitemOpen
  \bibfield  {author} {\bibinfo {author} {\bibfnamefont {W.}~\bibnamefont
  {Zhu}}, \bibinfo {author} {\bibfnamefont {X.}~\bibnamefont {Chen}}, \bibinfo
  {author} {\bibfnamefont {Y.-C.}\ \bibnamefont {He}}, \ and\ \bibinfo {author}
  {\bibfnamefont {W.}~\bibnamefont {Witczak-Krempa}},\ }\href {\doibase
  10.1126/sciadv.aat5535} {\  (\bibinfo {year} {2018}),\
  10.1126/sciadv.aat5535},\ \Eprint {http://arxiv.org/abs/1801.06177}
  {arXiv:1801.06177 [cond-mat.str-el]} \BibitemShut {NoStop}%
\bibitem [{\citenamefont {Villain}(1975)}]{Villain:1974ir}%
  \BibitemOpen
  \bibfield  {author} {\bibinfo {author} {\bibfnamefont {J.}~\bibnamefont
  {Villain}},\ }\href {\doibase 10.1051/jphys:01975003606058100} {\bibfield
  {journal} {\bibinfo  {journal} {J. Phys.(France)}\ }\textbf {\bibinfo
  {volume} {36}},\ \bibinfo {pages} {581} (\bibinfo {year} {1975})}\BibitemShut
  {NoStop}%
%%CITATION = JOPQA,36,581;%%
\bibitem [{\citenamefont {DeGrand}\ and\ \citenamefont
  {Toussaint}(1980)}]{DeGrand:1980eq}%
  \BibitemOpen
  \bibfield  {author} {\bibinfo {author} {\bibfnamefont {T.~A.}\ \bibnamefont
  {DeGrand}}\ and\ \bibinfo {author} {\bibfnamefont {D.}~\bibnamefont
  {Toussaint}},\ }\href {\doibase 10.1103/PhysRevD.22.2478} {\bibfield
  {journal} {\bibinfo  {journal} {Phys. Rev.}\ }\textbf {\bibinfo {volume}
  {D22}},\ \bibinfo {pages} {2478} (\bibinfo {year} {1980})},\ \bibinfo {note}
  {[,194(1980)]}\BibitemShut {NoStop}%
%%CITATION = PHRVA,D22,2478;%%
\bibitem [{\citenamefont {Murthy}\ and\ \citenamefont
  {Sachdev}(1990)}]{Murthy:1989ps}%
  \BibitemOpen
  \bibfield  {author} {\bibinfo {author} {\bibfnamefont {G.}~\bibnamefont
  {Murthy}}\ and\ \bibinfo {author} {\bibfnamefont {S.}~\bibnamefont
  {Sachdev}},\ }\href {\doibase 10.1016/0550-3213(90)90670-9} {\bibfield
  {journal} {\bibinfo  {journal} {Nucl. Phys.}\ }\textbf {\bibinfo {volume}
  {B344}},\ \bibinfo {pages} {557} (\bibinfo {year} {1990})}\BibitemShut
  {NoStop}%
%%CITATION = NUPHA,B344,557;%%
\bibitem [{\citenamefont {Pufu}\ and\ \citenamefont
  {Sachdev}(2013)}]{Pufu:2013eda}%
  \BibitemOpen
  \bibfield  {author} {\bibinfo {author} {\bibfnamefont {S.~S.}\ \bibnamefont
  {Pufu}}\ and\ \bibinfo {author} {\bibfnamefont {S.}~\bibnamefont {Sachdev}},\
  }\href {\doibase 10.1007/JHEP09(2013)127} {\bibfield  {journal} {\bibinfo
  {journal} {JHEP}\ }\textbf {\bibinfo {volume} {09}},\ \bibinfo {pages} {127}
  (\bibinfo {year} {2013})},\ \Eprint {http://arxiv.org/abs/1303.3006}
  {arXiv:1303.3006 [hep-th]} \BibitemShut {NoStop}%
%%CITATION = ARXIV:1303.3006;%%
\bibitem [{\citenamefont {Karthik}(2018)}]{Karthik:2018rcg}%
  \BibitemOpen
  \bibfield  {author} {\bibinfo {author} {\bibfnamefont {N.}~\bibnamefont
  {Karthik}},\ }\href {\doibase 10.1103/PhysRevD.98.074513} {\bibfield
  {journal} {\bibinfo  {journal} {Phys. Rev.}\ }\textbf {\bibinfo {volume}
  {D98}},\ \bibinfo {pages} {074513} (\bibinfo {year} {2018})},\ \Eprint
  {http://arxiv.org/abs/1808.08970} {arXiv:1808.08970 [cond-mat.str-el]}
  \BibitemShut {NoStop}%
%%CITATION = ARXIV:1808.08970;%%
\bibitem [{\citenamefont {Dyer}\ \emph {et~al.}(2013)\citenamefont {Dyer},
  \citenamefont {Mezei},\ and\ \citenamefont {Pufu}}]{Dyer:2013fja}%
  \BibitemOpen
  \bibfield  {author} {\bibinfo {author} {\bibfnamefont {E.}~\bibnamefont
  {Dyer}}, \bibinfo {author} {\bibfnamefont {M.}~\bibnamefont {Mezei}}, \ and\
  \bibinfo {author} {\bibfnamefont {S.~S.}\ \bibnamefont {Pufu}},\ }\href@noop
  {} {\  (\bibinfo {year} {2013})},\ \Eprint {http://arxiv.org/abs/1309.1160}
  {arXiv:1309.1160 [hep-th]} \BibitemShut {NoStop}%
%%CITATION = ARXIV:1309.1160;%%
\bibitem [{\citenamefont {Karthik}\ and\ \citenamefont
  {Narayanan}(2019{\natexlab{b}})}]{Karthik:2019mrr}%
  \BibitemOpen
  \bibfield  {author} {\bibinfo {author} {\bibfnamefont {N.}~\bibnamefont
  {Karthik}}\ and\ \bibinfo {author} {\bibfnamefont {R.}~\bibnamefont
  {Narayanan}},\ }\href {\doibase 10.1103/PhysRevD.100.054514} {\bibfield
  {journal} {\bibinfo  {journal} {Phys. Rev. D}\ }\textbf {\bibinfo {volume}
  {100}},\ \bibinfo {pages} {054514} (\bibinfo {year} {2019}{\natexlab{b}})},\
  \Eprint {http://arxiv.org/abs/1908.05500} {arXiv:1908.05500 [hep-lat]}
  \BibitemShut {NoStop}%
\end{thebibliography}%

\end{document}